\documentclass[12pt]{article}

\usepackage{amsmath}
\usepackage{latexsym}
\usepackage{amssymb}
\usepackage{verbatim}
\usepackage{graphicx}
\usepackage{color}

 \def\be{\begin{equation}}
 \def\ee{\end{equation}}
 \def\bea{\begin{eqnarray}}
 \def\eea{\end{eqnarray}}
 \def\bei{\begin{itemize}}
 \def\eei{\end{itemize}}
 \def\bs{\begin{slide}}
 \def\es{\end{slide}}
 \def\nn{\nonumber}

 \def\L{\mathcal{L}}
 
 \def\a{\alpha}
 
 \def\b{\beta}

 \def\d{\delta}
 \def\D{\Delta}
 \def\m{\mu}
 \def\n{\nu}
 \def\t{\tau}
 
 \def\th{\theta}

 \def\l{\lambda}

 \def\f{\phi}

 \def\MZO{M_{Z_0}}
 
 \def\({\left(}
 \def\){\right)}
 \def\[{\left[}
 \def\]{\right]}
 
 \def\gt{\tilde{g}}

 \def\vf{\varphi}
 \def\la{\langle}
 \def\ra{\rangle}
 \def\Tr{\textnormal{Tr}}

 \def\thth{\theta^2 \bar\theta^2}
 
 \def\QQ{{q_Q}}
 \def\QU{{q_{U^c}}}
 \def\QD{{q_{D^c}}}
 \def\QL{{q_L}}
 \def\QE{{q_{E^c}}}
 \def\QHu{{q_{H_u}}}
 \def\QHd{{q_{H_d}}}
 \def\cA{\mathcal{A}}

 \newcommand{\bth}{{\bf 3}}
 \newcommand{\btw}{{\bf 2}}
 \newcommand{\bon}{{\bf 1}}
 
 \def\ds#1{#1\kern-1ex\hbox{/}}
 \def\sla{\raise.15ex\hbox{$/$}\kern-.57em}
 \def\TeV{\text{ TeV}}
 \def\GeV{\text{ GeV}}
 \def\MVA{M_{V_A}}

\def\gt{\tilde{g}}

\def\ut{\tilde{u}}
\def\dt{\tilde{d}}
\def\ct{\tilde{c}}
\def\st{\tilde{s}}

\def\et{\tilde{e}}
\def\mt{\tilde{\mu}}

\def\nt{\tilde{\nu}}

\begin{document}

\begin{titlepage}

\rightline{ROM2F/2011/02}

\vskip 2cm

\centerline{{\large\bf Supersymmetry Breaking in a Minimal Anomalous Extension of the MSSM}}

\vskip 1cm

\centerline{A. Lionetto\footnote{Andrea.Lionetto@roma2.infn.it}$^\natural$ and
            A. Racioppi\footnote{Antonio.Racioppi@kbfi.ee}$^\flat$}

\vskip 1cm

\centerline{$^\natural$ Dipartimento di Fisica dell'Universit\`a di Roma ,
``Tor Vergata" and}
\centerline{I.N.F.N.~ -~ Sezione di Roma ~ ``Tor Vergata''}
\centerline{Via della Ricerca  Scientifica, 1 - 00133 ~ Roma,~ ITALY}
\vskip 0.25cm
\centerline{$^\flat$ National Institute of Chemical Physics and Biophysics,}
\centerline{Ravala 10, Tallinn 10143, Estonia}
\vskip 1cm

\begin{abstract}
We study a supersymmetry breaking mechanism in the context of a minimal
anomalous extension of the MSSM.
The anomaly cancellation mechanism is achieved through suitable counterterms in the effective action, i.e. Green-Schwarz
terms. We assume that the standard MSSM superpotential is perturbatively realized, i.e. all terms allowed by gauge symmetries, except for the $\mu$-term which has a non-perturbative origin. The presence of this term is expected in many intersecting D-brane models which can be considered as the ultraviolet completion of our model. We show how soft supersymmetry breaking terms arise in this framework and we study the effect of some phenomenological constraints on this scenario.
\end{abstract}

\end{titlepage}

\section{Introduction}
The LHC era has begun and the high energy physics community is
analyzing and discussing the first results.
One of the key goals of LHC, besides shedding light on the electroweak (EW) symmetry breaking sector of the Standard Model (SM), is to find some signature of physics beyond the SM.
Supersymmetric particles and extra neutral gauge bosons $Z'$ are widely studied examples of such signatures.
A large class of phenomenological and string models aiming to describe the low energy physics accessible to LHC predict the existence of additional abelian $U(1)$ gauge groups as well as $N=1$ supersymmetry softly broken roughly at the TeV scale.
In particular in string theory the presence of extra anomalous $U(1)$'s seems ubiquitous. D-brane models in orientifold vacua contain several abelian factors and they are typically anomalous~\cite{Pradisi:1988xd}-\cite{Kiritsis:2007zz}.
In~\cite{Anastasopoulos:2008jt} we studied a string inspired extension of the (Minimal Supersymmetric SM) MSSM with an additional anomalous $U(1)$
(see~\cite{corianoU1} for other anomalous $U(1)$ extensions of the SM and see~\cite{corianoU1susy} for extensions of the MSSM).
The term anomalous refers to the peculiar mechanism of gauge anomaly cancellation~\cite{GSmechanism} which does not rely on the fermion charges but rather on the presence of suitable counterterms in the effective action. These terms are usually dubbed as Green-Schwarz (GS)~\cite{corianoU1,Ibanez:1999it} and Generalized Chern-Simons (GCS)~\cite{de Wit:1984px}-\cite{DeRydt:2007vg}. They can be considered as the low energy remnants of the higher dimensional anomaly cancellation mechanism in string theory.
In our model we assumed the usual MSSM superpotential and soft supersymmetry breaking terms allowed by the symmetries (the well known result~\cite{Girardello:1981wz}). In this paper we address the question of the origin of the latter in the context of a global supersymmetry breaking mechanism. This means that we do not rely on a supergravity origin of the soft terms but rather on a local setup based for example on intersecting D-brane constructions in superstring theory in which gravity is essentially decoupled (see for instance~\cite{Fucito:2010dk} for a recent attempt in this direction). Moreover in~\cite{Anastasopoulos:2008jt} we made the assumption that all the MSSM superpotential terms were perturbatively realized, i.e. allowed by the extra abelian $U(1)$ symmetries. In the following we assume instead that the $\mu$-term is perturbatively forbidden. The origin of this term is rather non-perturbative and can be associated to an exotic instanton contribution which naturally arises from euclidean D-brane in the framework of a type IIA intersecting brane model (see~\cite{Blumenhagen:2009qh} and references therein).

The paper is organized as follows: in Sec.~\ref{sec:setup} we describe the basic setup of the model and we discuss the perturbative and non-perturbative origin of the superpotential terms. We argue how the latter can naturally come from an intersecting D-brane model considered as the ultraviolet (UV) completion of our model. In Sec.~\ref{sec:susybreak} we describe the (global) supersymmetry breaking mechanism that gives mass to all the soft terms. In Sec.~\ref{sec:vectormassmatrix} we compute the gauge vector boson masses while in Sec.~\ref{sec:scalpotsec} we study the scalar potential of the theory in the neutral sector.
In Sec.~\ref{sec:neutralinos} we describe the neutralino sector while in Sec.~\ref{sec:sfermions} we describe the sfermion mass matrices.
In Sec.~\ref{sec:pheno} we study the phenomenology of our model and the bounds that can be put by some experimental constraints. Finally in Sec.~\ref{sec:end} we draw our conclusions.

\section{Model Setup}\label{sec:setup}
The model is an extension of the MSSM with two extra abelian gauge groups,
$U(1)_A$ and $U(1)_B$. The first one is anomalous while the second one is
anomaly free. This assumption is quite generic since in models with several anomalous $U(1)$ symmetries there exists a unique linear combination which is anomalous while the other combinations are anomaly free.
The charge assignment for the chiral superfields is shown in
Table~\ref{QTable}.
  \begin{table}[ht]
  \centering
  \begin{tabular}[h]{|c|c|c|c|c|c|}
   \hline             & $SU(3)_c$    & $SU(2)_L$ & $U(1)_Y$ & $U(1)_A$   &
$U(1)_B$      \\
   \hline   $Q_i$     & $\bth$       &  $\btw$   &  $1/6$   & $q_{Q}$   & 0
\\
   \hline   $U^c_i$   & $\bar \bth$  &  $\bon$   &  $-2/3$  & $q_{U^c}$ & 0
\\
   \hline   $D^c_i$   & $\bar \bth$  &  $\bon$   &  $1/3$   & $q_{D^c}$ & 0
\\
   \hline   $L_i$     & $\bon$       &  $\btw$   &  $-1/2$  & $q_{L}$   & 0
\\
   \hline   $E^c_i$   & $\bon$       &  $\bon$   &  $1$     & $q_{E^c}$ & 0
\\
   \hline   $H_u$     & $\bon$       &  $\btw$   &  $1/2$   & $q_{H_u}$ & 0
\\
   \hline   $H_d$     & $\bon$       &  $\btw$   &  $-1/2$  & $q_{H_d}$ & 0
\\
   \hline   $\Phi^+$    & $\bon$       &  $\bon$   &  0       & 1         & 1
\\
   \hline   $\Phi^-$    & $\bon$       &  $\bon$   &  0       & -1        & -1
 \\
   \hline
  \end{tabular}
  \caption{Charge assignment.}\label{QTable}
  \end{table}
The vector and matter chiral multiplets undergo the usual gauge transformations
\bea
V&\to&V+i\(\Lambda-\Lambda^\dagger\)\nonumber\\
\Phi&\to& e^{-iq\Lambda}\Phi
\eea
The anomaly cancellation of the $U(1)_A$ gauge group is achieved by the four dimensional analogue of
the higher dimensional GS mechanism which involves the St\"uckelberg superfield
$S=s+2\theta\psi_S+\theta^2 F_S$ transforming as a shift
\be
S\to S-2iM_{V_A}\Lambda
\label{stuckelberg}
\ee
where $M_{V_A}$ is a mass parameter related to the anomalous $U(1)_A$ gauge boson mass.
It turns out that not all the anomalies can be cancelled in this way. In particular the so called mixed anomalies between anomalous and non anomalous $U(1)$'s require the presence of trilinear GCS counterterms.
For further details about the anomaly cancellation mechanism see Appendix \ref{app:ano} (see also for instance~\cite{Anastasopoulos:2008jt} and~\cite{Anastasopoulos:2006cz}).
The effective superpotential of our model at the scale $E=M_{V_A}$ is given by
\be
W=W_{MSSM}+\lambda e^{-kS}H_u H_d+m\Phi^+\Phi^-
\label{newsuperpot}
\ee
where $W_{MSSM}$ is given by
\be
W_{MSSM}=y_u^{i j} Q_i U^c_j H_u - y_d^{i j} Q_i D^c_j H_d - y_e^{i j} L_i E^c_j
H_d
\ee
which is the usual MSSM superpotential without the $\mu$-term which is forbidden
for a generic choice of the charges $q_{H_u}$ and $q_{H_d}$. The second term
in~(\ref{newsuperpot}) is the only gauge invariant coupling allowed between the St\"uckelberg
superfield and the two Higgs fields. This is the only allowed coupling with matter fields
for a field transforming as~(\ref{stuckelberg}). We will argue later about how
non perturbative effects can generate such a term. The last term in~(\ref{newsuperpot})
is a mass term for $\Phi^\pm$ which are charged under both $U(1)_A$ and
$U(1)_B$. These fields have been considered as supersymmetry breaking mediators
in the context of anomalous models by Dvali and Pomarol~\cite{Dvali:1996rj}.
They play a key role in generating gaugino masses.
In the effective lagrangian, besides the usual kinetic terms (they are charged under both $U(1)_A$ and $U(1)_B$),
the two $U(1)_B$ fields $\Phi^\pm$ couple to the gauge field strength
$W_a^\alpha$ through the dimension six effective operator
\be
{\cal{L}}_g=c_a \frac{\Phi^+\Phi^-}{\Lambda^2}W_a^\alpha W_{\alpha\,a}
\label{gauginooperator}
\ee
where $a=A,B,Y,2,3$, $\Lambda$ is the cut-off scale of the theory while $c_a$
are constants that have to be computed in the UV completion of the theory.

The non perturbative term in~(\ref{newsuperpot}) is expected to be generated in the effective action of
intersecting D-brane models which can be considered as the UV completion of our model.
This is the leading order term when the coupling
$H_u H_d$ is not allowed by gauge invariance.
In string theory there are many axions related to the GS mechanism of anomaly cancellation which are
charged under some Ramond-Ramond (RR) form. For example in type IIA orientifold model with
D6-branes, axion fields are associated to the $C_3$ RR-form (see for a recent review~\cite{Blumenhagen:2006ci}).
Instantons charged under this RR-form, such as euclidean E2-branes wrapping some $\gamma_3$
3-cycle in the Calabi-Yau (CY) compactification manifold, give a contribution to
the holomorphic couplings in the $N=1$ superpotential.
Our analysis does not rely on any concrete intersecting brane model but rather on the generic appearance of such instanton induced terms.
The exponential suppression factor of the classical instanton action is
\be
e^{-{\rm Vol_{E2}}/g_s}
\ee
where ${\rm Vol_{E2}}$ is the volume of the 3-cycle in the CY wrapped by a
$E2$-brane measured in string units while $g_s$ is the string coupling.
Such exponential factor is independent from the $d=4$ gauge coupling and thus
this instanton is usually termed as stringy or exotic instanton (see~\cite{Blumenhagen:2009qh} and~\cite{Bianchi:2009ij} and references therein). Moreover the instanton contribution can be sizable even in the case
$g_s\ll 1$ if ${\rm Vol_{E2}}\ll 1$ measured in string units.

In type IIA orientifold models with intersecting branes the complexified moduli,
whose imaginary part are the generalized axion fields (depending on the cycle $\gamma_3^i$),
can be written as
\be
U_i=e^{-\varphi}\int_{\gamma_3^i}\Omega_3+i\int_{\gamma_3^i} C_3
\ee
where $\varphi$ is the dilaton, $\Omega_3$ is the CY volume 3-form (which is a
complex form) and $C_3$ is the RR-form. The integral of this form is dual to the axion
whose shift symmetry is gauged in the GS mechanism.
The generic contribution of an $E2$ instanton is formally given by
\be
W\sim \prod_{i=1}^n \Phi_{a_i,\,b_i} e^{-S_{E2}}
\ee
where $\Phi_{a_i,\,b_i}$ are chiral superfields localized at the intersection of
two D6-branes described by open strings while $S_{E2}$ denotes the instanton
classical action
\be
e^{-S_{E2}}=\exp\left[-\frac{2\pi}{l_s^3}\(\frac{1}{g_s}\int_\gamma
Re(\Omega_3)-i\int_\gamma C_3\)\right]
\ee
This result can be immediately extended to the supersymmetric case which involves
the complete St\"uckelberg multiplet.
The appearance of the exponential suppression factor is dictated by the fact
that the superpotential is a holomorphic quantity. Thus the only allowed
functional dependence on the string coupling $g_s=e^{<\varphi>}$ and the axionic
field is an exponential.
Any other dependence can be excluded due to the shift
transformation~(\ref{stuckelberg}).

\section{Supersymmetry Breaking}\label{sec:susybreak}
The D-term contribution of the $U(1)_A$ vector multiplet $V_A$ relevant to
supersymmetry breaking
is given, in the limit of vanishing kinetic mixing $\delta_{YA},\delta_{AB}=0$, by the following lagrangian:
\be
\L=\frac{1}{2}D_A D_A+\sum_i g_A q_i\phi_i^\dagger D_A \phi_i+\xi
D_A
\label{eq:dterms}
\ee
where the sum is extended to all the scalars charged under the $U(1)_A$.
There is no D-term contribution related to the $U(1)_B$ except that of $\f^\pm$ since all the MSSM chiral fields are uncharged under $U(1)_B$ (see Table~\ref{QTable}).
The last term in~(\ref{eq:dterms}) is a tree-level field dependent Fayet-Iliopoulos (FI) term
which comes from the supersymmetrized St\"uckelberg lagrangian
\bea
\L_{axion} &=& \frac{1}{4} \left. \( S + S^\dagger + 2 M_{V_A} V_A \)^2
\right|_{\thth}
+\ldots\nonumber\\
&=& M_{V_A}\left.(S + S^\dagger)V_A\right|_{\thth}+\ldots\nonumber\\
&=& M_{V_A}\alpha D_A+\ldots
\eea
where in the last line $\alpha$ denotes the real part of the lowest component of
the St\"uckelberg chiral multiplet $s=\alpha+i\vf$. The fields $\alpha$ and $\vf$ are called the
saxion and the axion respectively\footnote{with a slight abuse of notation with respect to the previous section where we denoted the dilaton with $\varphi$.}.
We assume that the real part $\a$ gets an expectation value. This gives a
contribution to the gauge coupling constants which can be absorbed in the
following redefinition
  \be
    \frac{1}{16 g_a^2 \t_a}=\frac{1}{16 \gt_a^2 \t_a} - \frac{1}{2} b^{aa} \langle \a\rangle \label{couplingconst}
   \ee
where the gauge factors $\t_a$ take the values $1,1,1,1/2,1/2$ and the $b^{aa}$ constants are given in (\ref{bs}).
The tree-level FI term is then given by \be
\xi=M_{V_A}\left<\alpha\right> \ee Moreover in the following we
assume that 1-loop FI terms are absent (see the discussion
in~\cite{Poppitz:1998dj}). The FI term induces a mass term for the
scalars. This can be seen by solving the equations of motion for $D_A$
\be
 D_A+\sum_i g_A q_i\phi_i^\dagger  \phi_i+\xi=0
\ee
where the index $i$ runs over all chiral superfields. The D-term
contribution to the scalar potential is given by \be
V(\phi_i,\phi_i^\dagger)= \frac{1}{2}\(\xi+g_A\sum_i q_i
\left|\phi_i\right|^2\)^2 \ee The quadratic part gives the
scalar mass term \be \sum_i \xi g_A q_i
\left|\phi_i\right|^2=\sum_i m_i^2 \left|\phi_i\right|^2 \ee where
we have defined \be
 m_i^2=\xi g_A q_i=\left<\alpha\right> g_A M_{V_A} q_i= q_i m_\xi^2
\label{eq:dtermshift}
\ee
with
\be
m_\xi^2=\left<\alpha\right>g_A \MVA =g_A\xi
\ee
The typical scale for the mass $m_\xi$ is of the order of few hundreds of GeV if
$\MVA\sim\left<\alpha\right>\sim 1$ TeV and $g_A\sim 0.1$.
It is interesting to note that in this scenario a low subTeV supersymmetry
breaking scale $m_\xi$ is due to the St\"uckelberg mechanism which gives mass to $V_A$.
This is the most important difference with the scenario proposed
in~\cite{Dvali:1996rj}, where the scale $m_\xi$ is dynamically generated by some dynamics in a strong coupling regime.

Mass terms for the gauginos, i.e. $\lambda_a\lambda_{a}$, are generated by the dimension six effective operator~(\ref{gauginooperator}) in the broken phase where $\phi^\pm$
get vacuum expectation value (vev).
The contribution coming from this mechanism is
\be
 M_a =c_a \frac{\la F^+ \f^- \ra + \la F^- \f^+ \ra}{\Lambda^2} =c_a \frac{m \(v_+^2+v_-^2\)}{2\Lambda^2} \label{eq:gauginomass}
\ee
where $v_\pm/\sqrt2=\left<\phi_\pm\right>$ and where in the right hand side we have used the F-term equations of motion
for $F^\pm$
\be
F^\pm=-\frac{\partial W^*}{\partial \phi^{\pm*}}=-m\phi^{\mp*}
\ee
having assumed $m$ real without any loss in generality.
We assume $c_a=c$ for each $a$. This is an assumption of universality as a boundary condition at the cutoff scale $\Lambda$ which does not affect in a crucial way our analysis.
In section~\ref{sec:scalpotsec} we study the scalar potential of our model and we derive the conditions for having a vev for $\phi^{\pm}$ different from zero.
Since we are breaking supersymmetry in the global limit in which the Planck mass $M_P\to\infty$ the F-term induced contribution to the scalar masses
\be
m^2_i\sim \frac{\left< F_{\pm}\right>}{M_P^2}
\ee
vanishes leaving~(\ref{eq:dtermshift}) as the leading contribution.

The requirement of gauge invariance of the superpotential implies the following constraints on the $U(1)_A$ charges
  \bea
   q_{U^c} &=& - q_Q - q_{H_u}      \nn\\
   q_{D^c} &=& - q_Q - q_{H_d}  \nn\\
   q_{E^c} &=& - q_L - q_{H_d}    \label{QconstW}
  \eea
and
\be
 k = \frac{\QHu+\QHd}{2 \MVA} \label{keq}
\ee
As we said at the beginning of this section we assume that the net kinetic mixing between $U(1)_Y$ and $U(1)_A$ vanishes~\footnote{We postpone the discussion about the kinetic mixing between $U(1)_A$ and $U(1)_B$ to the next section.}.
There are two contributions for the $U(1)_Y-U(1)_A$ kinetic mixing: the 1-loop mixing $\delta_{YA}$ and $b^{YA}$ coming from
the GS coupling $S W_Y W_A$ (see eq.~(\ref{Laxion})). The following conditions imply a bound on the charges
 \bea
 \delta_{YA}=0  & \Rightarrow & \sum_f q_f  Y_f =0 \nn\\
 b^{YA}=0 & \Rightarrow & \sum_f q_f^2 Y_f=0
\label{QconstKeq}
\eea

where the sum is extended over all the chiral fermions in the theory. The constraints~(\ref{QconstKeq}) can be solved in terms of $q_Q$ and $q_L$. By using the conditions~(\ref{QconstW}) we get
\bea
 \QL &=& \frac{1}{4} \(3 \QHu - 4 \QHd \) \nn\\
 \QQ &=& -\frac{1}{12} \(5 \QHu - 2 \QHd \)
\label{QconstK}
\eea
The positive squared mass condition for the sfermions
\be
 m^2_{\tilde f} = g_A q_f \MVA \la \a \ra > 0
\ee
implies $q_f>0$ for all the sfermions having assumed without loss of generality $\la \a \ra > 0$.
Using the constraints (\ref{QconstW}) and (\ref{QconstK}) we get the allowed parameter space
\be
 \QHu<0 \, , \quad \frac{5}{2} \QHu < \QHd < \frac{3}{4} \QHu \label{sfermmassbounds}
\ee

\section{Scalar Potential}\label{sec:scalpotsec}
The key ingredient in our model is the instanton induced term in~(\ref{newsuperpot}) which couples the St\"uckelberg field to the Higgs fields. The $\theta^2$ component of this superpotential term gives the following contribution to the lagrangian
\bea
W_{inst}|_{\theta^2}& =&\lambda e^{-kS}H_u H_d|_{\theta^2}\nonumber\\
&=& \l e^{-k s} h_u F_d + \l e^{-k s} F_u h_d - \l k e^{-k s} F_S h_u h_d +\nn\\
    && \sqrt2 \l e^{-k s} k \( h_u \psi_S \tilde h_d + h_d \psi_S \tilde h_u\) - \l e^{-k s} k^2 h_u h_d \psi_S \psi_S
\label{WHuHd}
\eea
where $F_{u,d}$ are the F-terms of $H_{u,d}$.
Solving the F-terms equations for $H_u$ and $H_d$ we get the following contributions for the instanton induced term in the scalar potential
\be
 V_{inst} = 2 \l^2 e^{-2 k \a } h_u^\dag h_u + 2 \l^2 e^{-2 k \a } h_d^\dag h_d + \l k e^{-k \a} \( e^{-i k \vf} F_S h_u h_d + h.c. \)
\ee
In the following we assume that $\alpha$ gets a vev different from zero and that the mass of this field is much higher than $\Lambda$ so that its dynamics is not described by the low energy effective action. From the point of view of the UV completion (for example a type IIA intersecting brane model) this amounts to saying that the closed string modulus related to $\alpha$ is stabilized. Moreover we made the assumption that the same dynamics that stabilizes $\a$ also fixes $F_S$.
By supersymmetry the saxion field $\a$, being part of the Stuckelberg multiplet, has a tree-level mass $M_{V_A}$.
Thus if we want to consider a frozen dynamics for $\a$ at the TeV scale we have to assume a mass parameter for the anomalous $U(1)_A$ just slightly above the TeV scale, i.e. $M_{V_A}> 1$ TeV.
In this way the effective instanton induced potential at a scale $E \simeq 1$ TeV is thus given by
\be
 V_{inst} = 2 \l^2 e^{-2 k \la \a \ra } h_u^\dag h_u + 2 \l^2 e^{-2 k \la \a \ra } h_d^\dag h_d  +
   \l k e^{-k \la \a \ra} \( \la F_S \ra e^{-i k \vf}  h_u h_d + h.c. \)
\ee
The first two terms are $\mu$-terms while the third one is a b-term.
The complete effective scalar potential is given by
\bea
 V &=& (|\m|^2 + m^2_{h_u}) \( |h_u^0|^2 + |h_u^+|^2 \) +
       (|\m|^2 + m^2_{h_d}) \( |h_d^0|^2 + |h_d^-|^2 \)\nn\\
    && + (|m|^2 + m^2_{\f^+}) |\f^+|^2 + (|m|^2 + m^2_{\f^-}) |\f^-|^2)\nn\\
    && + \[ b e^{-i k \vf} \( h_u^+ h_d^- - h_u^0 h_d^0 \) + h.c. \] \nn\\
    && + \frac{1}{8} (g_2^2 + g_Y^2) \( |h_u^0|^2 + |h_u^+|^2 - |h_d^0|^2 - |h_d^-|^2 \)^2
       + \frac{1}{2} g_2^2 \left| h_u^+ h_d^{0*} + h_u^0 h_d^{-*} \right|^2 \nn\\
    &&+ \frac{1}{2} g_A^2 \[ \QHu \( |h_u^0|^2 + |h_u^+|^2 \) + \QHd \( |h_d^0|^2 + |h_d^-|^2 \) + |\f^+|^2 - |\f^-|^2 \]^2 \nn\\
    &&+ \frac{1}{2} g_B^2 \[ |\f^+|^2 - |\f^-|^2 \]^2
 \label{Vhiggstotal}
\eea
where
\bea
 \mu &=& \sqrt2 \l e^{-k \la \a \ra } \label{mueq} \\
 b &=& \l k e^{-k \la \a \ra} \la F_S \ra \label{beq}
\eea
These relations give a solution of the well known $\mu$-problem since both terms have a common origin (see the analysis in Sec.~(\ref{fixpar})).
The soft squared masses are generated by the FI $U(1)_A$ term
\bea
 m^2_{h_u} &=& q_{H_u} m_\xi^2 \\
 m^2_{h_d} &=& q_{H_d} m_\xi^2 \\
 m^2_{\f^+} &=& m_\xi^2 \label{eq:mphiplusoft}\\
 m^2_{\f^-} &=& - m_\xi^2
\eea
with $m_\xi^2$ given by~(\ref{eq:dtermshift}).
The scalar potential depends on the following new parameters: $\la \a \ra$, $\la F_S \ra$, $\l$, $m$, $g_{A,B}$, $q_{H_{u,d}}$, $\MVA$.

In order to have a vacuum preserving the electromagnetism the charged field vevs must vanish. Thus we are left with the problem of finding a minimum for the neutral scalar potential
\bea
 V_0 &=& (|\m|^2 + m^2_{h_u}) |h_u^0|^2 + (|\m|^2 + m^2_{h_d}) |h_d^0|^2 - (b \, e^{-i k \vf} \, h_u^0 h_d^0 + h.c.) \label{Vhiggs} \\
     &&+ (|m|^2 + m^2_{\f^+}) |\f^+|^2 + (|m|^2 + m^2_{\f^-}) |\f^-|^2  \nn\\
    && + \frac{1}{8} (g_2^2 + g_Y^2) \( |h_u^0|^2 - |h_d^0|^2 \)^2\nn\\
    && + \frac{1}{2} g_A^2 \( q_{H_u} |h_u^0|^2 +q_{H_d}  |h_d^0|^2 + |\f^+|^2 - |\f^-|^2 \)^2 \nn\\
    &&+ \frac{1}{2} g_B^2 \[ |\f^+|^2 - |\f^-|^2 \]^2 \nn
\eea
Since there are no D-flat directions along which the quartic part vanishes, the potential is always bounded from below.
To find the minimum we solve $\partial V_0/\partial z^i=0$ where the scalar field $z^i$ runs over
$\{\vf,h_u^0,h_d^0,\f^+,\f^-\}$.
The conditions for having a non-trivial minimum boils down to the same condition of the MSSM
\be
 b^2 > (|\m|^2 + m^2_{h_u} )(|\m|^2 + m^2_{h_d})  \label{eq:destabilizeorigin}
\ee
Moreover in order to generate a mass term for the gauginos (see eq.~(\ref{eq:gauginomass})) the condition $v_-\neq 0$ must hold since $v_+=0$ due to the positive sign of the coefficient of the $\f^+$ quadratic term in ~(\ref{eq:mphiplusoft}).
This implies the following condition for the coefficient of the $\f^-$ quadratic term
\be
 |m|^2 + m^2_{\f^-} <0
\ee
The minimum is attained at $\vf=\f^+ =0$.
Actually since the potential for the axion $\vf$ is periodic the minimum condition holds for $\vf=2 n \pi/k$ with $n \in \mathbb{Z}$. All these minima are physically equivalent and thus we arbitrarily choose $n=0$.
The remaining three conditions imply the following constraints on the parameters
\bea
  m_{h_d}^2 + \mu^2 -b \, t_\b + \frac{1}{8} (g_Y^2+g_2^2) v^2 c_{2\b} + \frac{1}{2} g_A^2 \QHd \[v^2 \( \QHd c_\b^2+ \QHu s_\b^2 \) - v_-^2 \]  &=& 0
\label{minima1}\\
  m_{h_u}^2 + \mu^2 -b \, t_\b^{-1} - \frac{1}{8} (g_Y^2+g_2^2) v^2 c_{2\b} + \frac{1}{2} g_A^2 \QHu \[v^2 \( \QHd c_\b^2+ \QHu s_\b^2 \) - v_-^2 \]   &=& 0
\label{minima2}\\
\(g_A^2+g_B^2\)v_-^2- g_A^2 v^2 \( \QHd c_\b^2+ \QHu s_\b^2 \)+ 2 \( |m|^2 + m^2_{\f^-} \) &=& 0 \label{minima3}
\eea
where we have defined in order to keep a compact notation
\be
 c_\b = \cos\b, \quad  s_\b = \sin\b, \quad t_\b = \tan\b, \quad c_{2\b} = \cos(2\b), \quad s_{2\b} = \sin(2\b)
\ee
and as usual as $\tan\beta=v_u/v_d$.

In the previous discussion we treated the scalar potential in an exact way. In the following we want to introduce some useful approximation in order to compute the mass eigenstates.
Let us go back to the minima equations (\ref{minima1}-\ref{minima3}).
Supposing
\be
  v \ll  v_-
\label{vevhierarchyconditioneq}
\ee
we can neglect all the $g_A v$ terms.
With this approximation the minima equations read
\bea
  \tilde m_{h_d}^2 + \mu^2 -b \, t_\b + \frac{1}{8} (g_Y^2+g_2^2) v^2 c_{2\b}  &=& 0 \label{minima1approx} \\
  \tilde m_{h_u}^2 + \mu^2 -b \, t_\b^{-1} - \frac{1}{8} (g_Y^2+g_2^2) v^2 c_{2\b}    &=& 0 \label{minima2approx} \\
\(g_A^2+g_B^2\)v_-^2 +  2 \( |m|^2 + m^2_{\f^-} \) &=& 0 \label{minima3approx}
\eea
where we have defined
\bea
 \tilde m_{h_d}^2 &=& m_{h_d}^2 - \frac{1}{2} g_A^2 \QHd  v_-^2\label{mHd}\\
 \tilde m_{h_u}^2 &=& m_{h_u}^2 - \frac{1}{2} g_A^2 \QHu  v_-^2\label{mHu}
\eea
Equations (\ref{minima1approx}) and (\ref{minima2approx}) have the same functional form as in the MSSM case.
Moreover $v_-$ does not depend on any parameter of the visible sector.
Within this approximation the dynamics of the fields $\f^\pm$ is decoupled from that of the Higgs sector and thus the Higgs potential can be studied by fixing $\f^\pm$ at their vevs.
We get
\bea
 V &\simeq& (|\m|^2 + m^2_{h_u}) \( |h_u^0|^2 + |h_u^+|^2 \) +
       (|\m|^2 + m^2_{h_d}) \( |h_d^0|^2 + |h_d^-|^2 \) \nn \\
    && + \[ b e^{-i k \vf} \( h_u^+ h_d^- - h_u^0 h_d^0 \) + h.c. \] \nn\\
    && + {1\over 8} (g_2^2 + g_Y^2) \( |h_u^0|^2 + |h_u^+|^2 - |h_d^0|^2 - |h_d^-|^2 \)^2
       + {1\over 2} g_2^2 \left| h_u^+ h_d^{0*} + h_u^0 h_d^{-*} \right|^2 \nn\\
    &&+ {1\over 2} g_A^2 \[ \QHu \( |h_u^0|^2 + |h_u^+|^2 \) + \QHd \( |h_d^0|^2 + |h_d^-|^2 \) - \frac{1}{2} v_-^2 \]^2
\label{Vhapproxstep}
\eea
neglecting further constant terms in $v_-$. Close to the minima the relevant term in the last line of eq. (\ref{Vhapproxstep})
is the double product of the Higgs part with the $v_-^2$ term. Hence by using~(\ref{vevhierarchyconditioneq}) we finally get
\bea
 V_{h,\vf} &\simeq& (|\m|^2 + \tilde m^2_{h_u}) \( |h_u^0|^2 + |h_u^+|^2 \) +
       (|\m|^2 +\tilde  m^2_{h_d}) \( |h_d^0|^2 + |h_d^-|^2 \) \label{Vhapprox} \\
    && + \[ b e^{-i k \vf} \( h_u^+ h_d^- - h_u^0 h_d^0 \) + h.c. \] \nn\\
    && + {1\over 8} (g_2^2 + g_Y^2) \( |h_u^0|^2 + |h_u^+|^2 - |h_d^0|^2 - |h_d^-|^2 \)^2
       + {1\over 2} g_2^2 \left| h_u^+ h_d^{0*} + h_u^0 h_d^{-*} \right|^2 \nn
\eea
This potential has the same form (except for the contribution of the exponential term in $\varphi$) of the MSSM
potential and the corresponding minima equations are exactly given in eqs (\ref{minima1approx}) and (\ref{minima2approx}).
Thus all the well known MSSM results apply here \cite{Martin:1997ns}.

In particular one of the constraints is $t_\b \gtrsim 1.2$ \cite{Martin:1997ns} which implies\footnote{The presence of the extra field $\varphi$ does not affect this result since the minima conditions are the same as the MSSM.} $\tilde m^2_{h_u} < \tilde m^2_{h_d}$. By using the equations~(\ref{mHd}) and~(\ref{mHu}) we get
\be
 g_A \QHu \( \MVA \la \a \ra - \frac{1}{2} g_A  v_-^2 \) < g_A \QHd \( \MVA \la \a \ra - \frac{1}{2} g_A  v_-^2 \) \label{QHboundtanbstep}
\ee
By assuming $M_{V_A}> 1$ TeV, $v_-$ in the TeV range, $g_A\sim O(0.1)$
the term between brackets is positive and we get the following constraint
\be
 \QHu < \QHd \label{QHboundtanb}
\ee
for the $U(1)_A$ Higgs charges.

\subsection{Higgs mass matrices}
We discuss the mass eigenvalues starting from the exact form of the scalar potential (\ref{Vhiggstotal}),
switching to the approximated expression (\ref{Vhapprox}) when needed.
In the neutral sector the singlet scalar $\f^+$ does not mix with any other scalar so it is a mass eigenstate with square mass
\be
 M_{\f^+}^2 = 2 |m|^2 \label{eq:phiplusmass}
\ee
The same holds for the imaginary part of $\f^-$ which becomes the longitudinal mode of the gauge vector $Z_2$.
The mass matrix for the real scalar fields $\{\vf,Im(h_u^0),Im(h_d^0)\}$ is given by
\be
{\cal M}_S^{(Im)}=\(
\begin{array}{ccc}
 b \, t_\b              &   \dots                  & \dots                  \\
 b                      &   b \, t_\b^{-1}         & \dots                  \\
 -b \, k \, v \, s_\b   &   -b \, k \, v \, c_\b   & b \, k^2 \, v^2 \, c_\b s_\b
\end{array}
\)
\ee
The determinant of this matrix is zero. Two eigenvalues are zero which correspond to the Goldstone modes of $Z_0$ and $Z_1$.
The physical massive state is an axi-higgs state with mass given by
\be
 M_{A^0}^2 = \frac{2b}{ s_{2\b}} \[1 -\frac{1}{16}  \frac{\(\QHu+\QHd\)^2 v^2}{\MVA^2} s_{2\b}^2 \] \label{mA0}
\ee
where we used the relation~(\ref{keq}).
The mass matrix for the real scalar fields $\{Re(h_u^0)$, $Re(h_d^0)$, $\f_R^-\equiv Re(\f^-)\}$ reads as
\bea
&&\!\!\!\!\!\!\!\!\!{\cal M}_S^{(Re)}=   \label{CPevenmatrix}\\
&&\!\!\!\!\!\!\! {\(
\begin{array}{ccc}
   \(\frac{1}{4} g_{EW}^2+g_A^2 \QHd^2 \)v^2 c_\b^2+b t_\b   &  \dots   & \dots \\
  \!\! -b-\(\frac{1}{4} g_{EW}^2 -g_A^2 \QHd \QHu \)v^2 c_\b s_\b &  \(\frac{1}{4} g_{EW}^2+g_A^2 \QHu^2 \)v^2 s_\b^2+b t_\b^{-1} &  \dots \\
  -g_A^2 \QHd v \, v_- c_\b                                       &  -g_A^2 \QHu v \, v_- s_\b                                         & \(g_A^2+g_B^2\) v_-^2 \\
\end{array}
\)
}\nn
\eea
where $g_{EW}^2=(g_Y^2+g_2^2)$. The matrix can be diagonalized exactly but the results are cumbersome and difficult to read.
It is much more convenient starting from the approximated potential (\ref{Vhapprox}) neglecting the mixing between Higgses and $\f^-$.
In this case we can apply the MSSM equations and get the following mass eigenvalues
\bea
 \!\!\!\!\!
 M^2_{h^0,H^0} &\simeq&\frac{1}{2} \( \frac{2b}{ s_{2\b}}  \mp  \sqrt{ \(\frac{2b}{ s_{2\b}}-\frac{1}{4} (g_Y^2+g_2^2) v^2\)^2+2b  (g_Y^2+g_2^2) v^2 s_{2\b} } \) \\
 \!\!\!\!\!
 M_{\f^-_R}^2 &\simeq& \(g_A^2+g_B^2\) v_-^2 \label{eq:mphiminusR}
\eea

The charged sector is unchanged with respect to the MSSM, so
\be
 M^2_{H^\pm} = \frac{2b}{s_{2\b}}+M_W^2
\ee

As in the standard MSSM case the mass of the lightest Higgs $M_{h^0}$ has a theoretical bound.
It is a well known problem in the MSSM that the upper bound \cite{Inoue:1982ej} is not compatible with the LEP bound \cite{Schael:2006cr}.
In our case the bound is increased due to the presence of $D_A$-term corrections
\be
 M_{h^0}^2 < \frac{1}{4} (g_Y^2+g_2^2) v^2 c^2_{2\b} + \frac{1}{4} g_A^2 v^2 \[ \QHd + \QHu + \( \QHd - \QHu \) c_{2\b} \]^2
\ee
where the first term is the MSSM bound.
In principle, for arbitrary high values of $g_A \QHd$, $g_A \QHu$ we get an increasing  upper bound.
However, as in the standard MSSM case, $M_{h^0}^2$ undergoes to relatively drastic quantum corrections~\cite{Martin:1997ns}.
Hence in Section \ref{sec:pheno} we consider tree-level masses for all the particles except for $h_0$ for which we use the 1-loop corrected expression (see eq. (\ref{eq:mh0loop})).

\section{Vector mass matrix} \label{sec:vectormassmatrix}
We now discuss the vector mass matrix.
All the neutral scalars could in principle take a vev different from zero, hence we assume
\bea
 \la \f_\pm \ra &=& \frac{v_\pm}{\sqrt2} \\
 \la h^0_{u,d} \ra &=& \frac{v_{u,d}}{\sqrt2}
\eea
%
The neutral vector square mass matrix in the base $(V_B,V_A,V_Y,V_2^3)$ is
\be {\cal M}_V=\( {
\begin{array}{cccc}
 g_B^2 v_\f^2      & \dots                                                          & \dots                       & \dots \\
 g_A g_B v_\f^2    & g_A^2 \[\( c_\b^2 \QHd^2 + s_\b^2 \QHu^2 \) v^2+v_\f^2\] +\MVA^2 & \dots                       & \dots \\
 0                & \frac{1}{2} g_A g_Y q_H(\b) v^2                                & \frac{1}{4} g_Y^2 v^2       & \dots \\
 0                & -\frac{1}{2} g_A g_2 q_H(\b) v^2                               & -\frac{1}{4} g_Y g_2 v^2    & \frac{1}{4} g_2^2 v^2
\end{array}
} \)
\ee
where
\bea
 v_\f^2 &=& v_+^2 + v_-^2 \\
 v^2 &=& v_u^2 + v_d^2 \\
 q_H(\b) &=& \(s_\b^2 \QHu -c_\b^2 \QHd\)
\eea
By taking $\MVA > 1$ TeV (see Sec. \ref{sec:scalpotsec}), $V_A$ can be considered as decoupled from the low energy gauge sector (namely $E\lesssim 1$ TeV),
and we can ignore with very good approximation any mixing term\footnote{
The kinetic mixing between $U(1)_A$ and $U(1)_B$ deserves some comment, in particular if we relax the $\MVA > 1$ TeV assumption.
Actually the presence of this mixing turns out to be irrelevant for the phenomenology of the visible sector.
Anyway one has to take into account that for $\Tr \( q_A q_B \) \neq 0$ such a mixing arises at the 1-loop level. In such a case it can be assumed that the two $U(1)$'s are in the kinetic diagonalized basis with $\Tr \( q_A q_B \) = 0$ thanks to
some additional heavy chiral multiplet charged under both $U(1)_A$ and $U(1)_B$. These multiplets generate a counterterm
in the effective theory that cancels against $\d_{AB}$ making the net kinetic mixing term equal to zero.
This mechanism is analogous to the anomaly cancellation one where the GS mechanism can be generated by an anomaly free theory with some heavy chiral fermion integrated out of the mass spectrum~\cite{Anastasopoulos:2006cz}.
} involving $V_A$. From now on we will apply this approximation.

Since $V_B$ is a hidden gauge boson, it is decoupled from the SM sector.
The charged vector sector is unchanged with respect to the MSSM, so
\bea
 W^\pm_\m &=& \frac{V_2^{1\m} \mp i V_2^{2\m}}{\sqrt2} \\
 M^2_W &=& \frac{1}{4} g_2^2 v^2
\eea

\section{Neutralinos} \label{sec:neutralinos}
In comparison with the standard MSSM we now have five new neutral fermionic fields: $\psi_S$, $\l_A$, $\l_B$, $\tilde \f^\pm$.
However {under the assumption $\MVA > 1$ TeV, $\psi_S$ and $\l_A$} are not in the low energy sector because of the $\MVA$ mass term\footnote{
We stress that the $\psi_S-\l_A$ sector presents a different parameters choice with respect to \cite{Fucito:2008ai}-\cite{Fucito:2010dj},
where we realized a scenario in which the mixing between $\psi_S$ and $\l_A$ was suppressed.}. Thus we have
    \be
    \L_{\mbox{neutralino mass}} = -\frac{1}{2} (\psi^{0})^T {\cal M}_{\tilde N} \psi^0 + h. c.
    \ee
where
   \be(\psi^{0})^T= (\l_B, \ \tilde \f^-, \ \tilde \f^+, \ \l_Y, \ \l_2^0,\ \tilde h_d^0,\ \tilde h_u^0) \label{neutrbase}
    \ee
In this basis the neutralino mass matrix $ {\cal M}_{\tilde N} $ is written as
\be
{\cal M}_{\tilde N}     =
{
 \(\begin{array}{ccccccccc}
    M_B      & \ldots   & \ldots &   \ldots           &   \ldots           & \ldots             & \ldots    \\
    -g_B v_- & 0        & \ldots &   \ldots           &   \ldots           & \ldots             & \ldots\\
    0        & -m       & 0      &   \ldots           &   \ldots           & \ldots             & \ldots\\
    0        & 0        & 0      &  M_1               &   \ldots           & \ldots             & \ldots\\
    0        & 0        & 0      & 0                  &   M_2              & \ldots             & \ldots\\
    0        & 0        & 0      & -\frac{g_1 v_d}{2} & \frac{g_2 v_d}{2}  & 0                  & \ldots  \\
    0        & 0        & 0      &  \frac{g_1 v_u}{2} & -\frac{g_2 v_u}{2} & -\m                & 0
 \end{array}\)}
 \label{Neutralinosmatrix}
\ee
where $\m$ is given in eq. (\ref{mueq}).
We remind that gaugino masses arise from the Dvali-Pomarol term~(\ref{gauginooperator}).

$ {\cal M}_{\tilde N} $ factorizes in a $4 \times 4$ MSSM block in the lower right corner, and in a $3 \times 3$ new sector block in the upper left corner.
The new sector block is given by the $\l_{B}$ and $\tilde \f^\pm$ contributions.
This last block has a MSSM-like structure that can be easily understood just considering the superpotential (\ref{newsuperpot}),
the gaugino masses (\ref{eq:gauginomass}) and by reminding that $\f^-$ gets a vev $v_-$ different from zero, while  $v_+=0$.

Finally there are also corrections coming from the anomalous axino couplings: F-term couplings of the type $b^{aa} \la F_S \ra \l_a \l_a$
and D-term couplings of the type $b^{aa} \psi_S \l_a \la D_a \ra$, and corrections coming from
the superpotential term $e^{-kS}H_u H_d+h.c.$. However such corrections are always subdominant and thus we neglect them with very good approximation.

We assume the
lightest supersymmetric particle (LSP) in our model comes from the neutralino sector. In Sec.~\ref{sec:pheno} we show the parameter regions in which this holds true. In order to ensure that the neutralino is the LSP we keep fixed the gravitino mass $m_{3/2}\sim {\rm O(TeV)}$ in the limit $M_P\to\infty$.

\section{Sfermion masses} \label{sec:sfermions}
The sfermion masses receive several contributions.
As we have seen in Sec.~\ref{sec:susybreak} the leading contribution comes from the induced soft masses~(\ref{eq:dtermshift}).
But there are further contributions. We have MSSM-like contributions: F-term corrections proportional to the Yukawa couplings,
 $D_Y$ and $D_2$ term correction from the Higgs sector. Moreover there are $D_A$ term corrections from the Higgs and $\f^-$ sector. As an aside, the appearance of such terms in the low-energy action, given our assumption $M_{V_A}> 1$ TeV, can be understood in terms of quantum corrections to Kahler potential~\cite{ArkaniHamed:1998nu}.
Considering the first two families we neglect the corresponding Yukawa couplings (the so called third family approximation). In this approximation the mass eigenvalues are given by
\bea
 m^2_{\ut_L} \simeq m^2_{\ct_L}  &=& m^2_{\tilde Q} + \( \frac{1}{3} g_Y^2 - g_2^2 \) \frac{\D v^2}{8}   + \QQ \tilde{m}^2_{D_A} \label{mutL}\\
 m^2_{\ut_R} \simeq m^2_{\ct_R}  &=& m^2_{\tilde U^c}     -g_Y^2                      \frac{\D v^2}{6}   + \QU \tilde{m}^2_{D_A} \label{mutR}\\
 m^2_{\dt_L} \simeq m^2_{\st_L}  &=& m^2_{\tilde Q} + \( \frac{1}{3} g_Y^2 + g_2^2 \) \frac{\D v^2}{8}   + \QQ \tilde{m}^2_{D_A}  \label{mdtL}\\
 m^2_{\dt_R} \simeq m^2_{\st_R}  &=& m^2_{\tilde D^c} +    g_Y^2                      \frac{\D v^2}{12}  + \QD \tilde{m}^2_{D_A} \label{mdtR}\\
 m^2_{\nt_e} = m^2_{\nt_\m} &=& m^2_{\tilde L} - \( g_Y^2 + g_2^2 \) \frac{\D v^2}{8} \label{mnetL} + \QL \tilde{m}^2_{D_A}\\
 m^2_{\et_L} \simeq m^2_{\mt_L}  &=& m^2_{\tilde L} - \( g_Y^2 - g_2^2 \) \frac{\D v^2}{8} \label{metL}  + \QL \tilde{m}^2_{D_A}\\
 m^2_{\et_R} \simeq m^2_{\mt_R}  &=& m^2_{\tilde E^c} +    g_Y^2          \frac{\D v^2}{4} \label{metR}  + \QE \tilde{m}^2_{D_A}
\eea
The first terms on the right hand side $m^2_{{\tilde Q},{\tilde U^c},{\tilde D^c},{\tilde L},{\tilde E^c}}$ are the corresponding soft masses (\ref{eq:dtermshift}),
the second terms are the $D_{Y,2}$ contributions with $\D v^2=v_u^2-v_d^2=-v^2 c_{2\b}$, while the last terms are the
$D_A$ corrections given by
\be
\tilde{m}^2_{D_A} = \frac{1}{2} \( \QHu v_u^2 + \QHd v_d^2 - v_-^2 \)
\ee
There is an approximated degeneracy between the sfermions with the same charges.

The mass matrix for the third family sfermions is parametrized as
\be
  {\cal M}_{\tilde f}^2=\left(\begin{array}{cc} { M^{\tilde f}_{LL} }^2  \,\, { M^{\tilde f}_{LR} }^2\\
                                                { M^{\tilde f}_{LR} }^2  \,\, { M^{\tilde f}_{RR} }^2  \end{array}\right)
\label{3rdsfmm}
\ee
where the off-diagonal terms are generated by F-term corrections proportional to the Yukawa couplings.
The stop mass matrix elements are
\bea
 &&{ M^{\tilde t}_{LL} }^2 = m_t^2 + m^2_{\tilde Q} + \( \frac{1}{3} g_Y^2 - g_2^2 \) \frac{\D v^2}{8}    + \QQ \tilde{m}^2_{D_A} \nn\\
 &&{ M^{\tilde t}_{RR} }^2 = m_t^2 + m^2_{\tilde U^c}    -g_Y^2                        \frac{\D v^2}{6}   + \QU \tilde{m}^2_{D_A} \nn\\
 &&{ M^{\tilde t}_{LR} }^2 = -\m  \, m_t \, t_\b^{-1}  \label{stopmassmatrixelements}
\eea
The sbottom mass matrix elements are
\bea
 &&{ M^{\tilde b}_{LL} }^2 =   m_b^2 + m^2_{\tilde Q} + \( \frac{1}{3} g_Y^2 + g_2^2 \) \frac{\D v^2}{8}    + \QQ \tilde{m}^2_{D_A} \nn\\
 &&{ M^{\tilde b}_{RR} }^2 =   m_b^2 + m^2_{\tilde D^c} +    g_Y^2                        \frac{\D v^2}{12} + \QD \tilde{m}^2_{D_A} \nn\\
 &&{ M^{\tilde b}_{LR} }^2 =  -\m \, m_b \, t_\b   \label{sbottommassmatrixelements}
\eea
The stau mass matrix elements are
\bea
 &&{ M^{\tilde \t}_{LL} }^2 =   m_\t^2 + m^2_{\tilde L} - \( g_Y^2 - g_2^2 \) \frac{\D v^2}{8}    + \QL \tilde{m}^2_{D_A} \nn\\
 &&{ M^{\tilde \t}_{RR} }^2 =   m_\t^2 + m^2_{\tilde E^c} +     g_Y^2 \frac{\D v^2}{4}            + \QE \tilde{m}^2_{D_A} \nn\\
 &&{ M^{\tilde \t}_{LR} }^2 =  -\m \, m_\t \, t_\b  \label{staumassmatrixelements}
\eea
The tau sneutrino mass is
\be
 m^2_{\nt_\t}=m^2_{\tilde L} -  \( g_Y^2 + g_2^2 \) \frac{\D v^2}{8}  + \QL \tilde{m}^2_{D_A} \label{mntautL}
\ee
where $m_t$, $m_b$ and $m_\t$ are the masses of the corresponding
standard fermions (i.e. further F-term contributions proportional
to the Yukawa couplings). The structure of the diagonal terms of
(\ref{3rdsfmm}) is the same as in eq.~(\ref{mutL})-(\ref{metR}):
soft masses, MSSM D-term contribution and $D_A$ term correction.
Furthermore we stress that there is a mass degeneracy between the
three sneutrinos $\tilde \n_{e,\m,\t}$ since the soft masses
(\ref{eq:dtermshift}) are flavor blind.

\section{Phenomenology}\label{sec:pheno}
In the following we derive the phenomenological consequences of our scenario.
Following our assumption of having a mass parameter for the anomalous $U(1)_A$ just slightly above the TeV scale, we fix $M_{V_A}=10$ TeV. The mass scale in the gaugino sector $\Lambda$ is set to be $O(M_{V_A})$.

\subsection{Charge Bounds} \label{sec:Charge Bounds}
The model parameter space can in principle be constrained by precision EW measurements~\cite{:2005ema}.
\begin{figure}[t]
\centering
\includegraphics[scale=0.3]{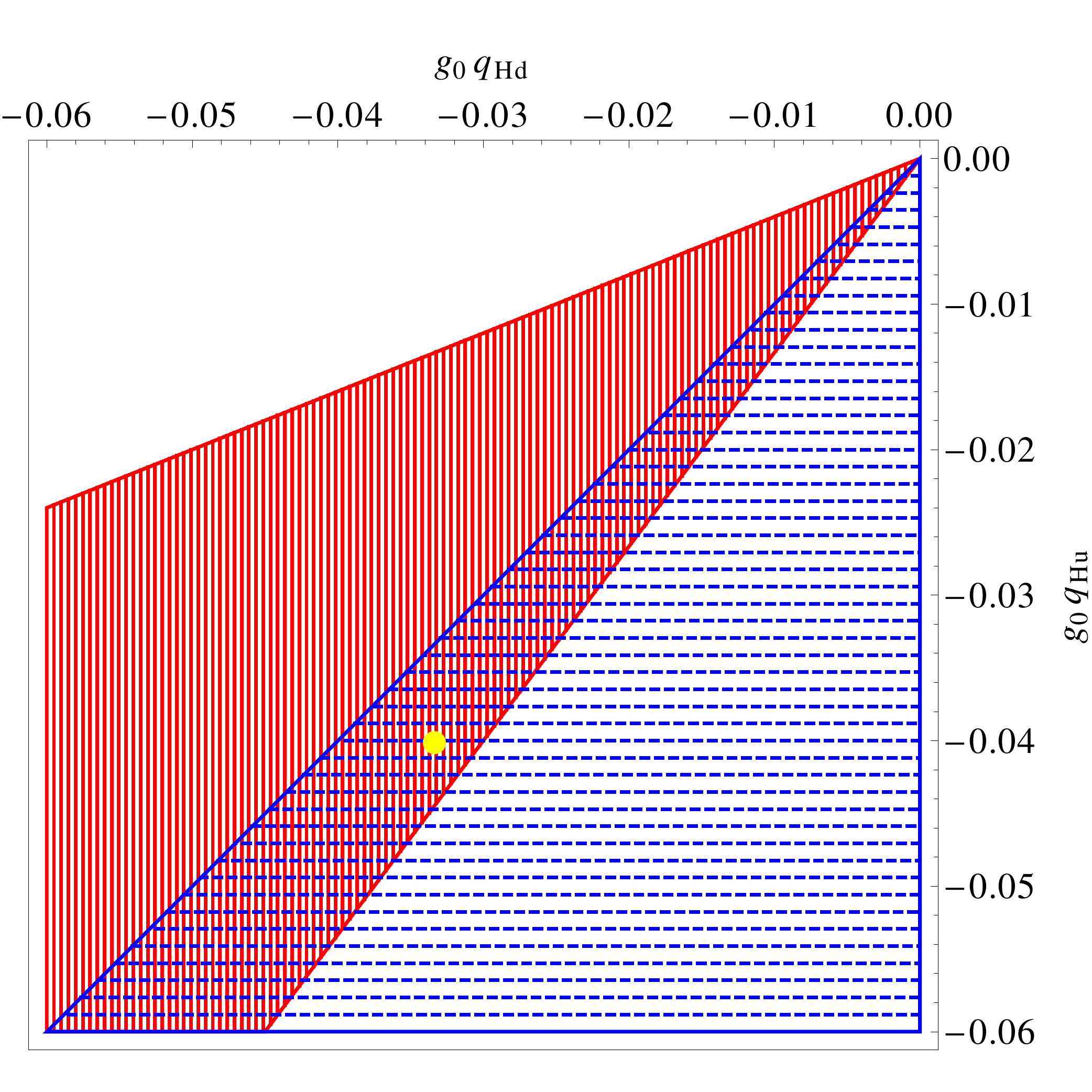}
\caption{Higgs couplings bounds. The yellow spot represents our charge choice.}
\label{fig:QHbound}
\end{figure}
However, since $\MVA=10$ TeV every value of $g_A \QHu$ and $g_A \QHd$ is allowed by EW precision data if $|g_A \QHu|,|g_A \QHd|\lesssim 0.1$.
So the only relevant constraints are ({\ref{sfermmassbounds}) and (\ref{QHboundtanb}), that are plotted respectively with a red and a blue region,
in Fig.~\ref{fig:QHbound} in the plane ($g_A \QHu$, $g_A \QHd$).
}

\subsection{Free Parameters} \label{fixpar}

Here we discuss which parameters remain free in our model after all the constraints discussed in the previous sections are imposed.
Our choice for the Higgs $U(1)_A$ charges corresponds to the yellow spot in Fig.~\ref{fig:QHbound}
\bea
 &&g_A = 0.1 \qquad \MVA = 10 \TeV \nn\\
 &&\QHd = -(1/3) \qquad \QHu = -(2/5) \label{commonnumeric}
\eea

In order to fix the remaining parameters ($\la \a \ra$, $\la F_S \ra$, $\l$, $m$, $g_B$) we assume $v\simeq 246$ GeV
and then we choose some benchmark value for $g_B$ and $v_-$ in the $U(1)_B$ sector\footnote{We remind that $v_+=0$ (see Section~(\ref{sec:scalpotsec})).}:
\bea
 &&A) \quad g_B=0.4 \qquad v_-=5 \TeV \label{caseA}\\
 &&B) \quad g_B=0.1 \qquad v_-=4 \TeV \label{caseB}
\eea
The next step is to solve the minima conditions (\ref{minima1})-(\ref{minima3}) determining $\la F_S \ra$, $\l$, $m$ as function of $\la \a \ra$.
In the limit in which $v^2\ll\MVA \la \a \ra,v_-^2$, we get

\bea
 \l^2 &\simeq& \frac{1}{8}\,  e^{\frac{2 \la \a \ra g_A (\QHd+\QHu)}{\MVA}}
 \Big[ g_A \(  g_A    v_-^2-2 \la \a \ra \MVA \) (\sec (2 \beta )    (\QHd-\QHu)+\QHd+\QHu)\Big]\nn\\
 \la F_S \ra &\simeq& -\, e^{\frac{\la \a \ra g_A (\QHd+\QHu)}{\MVA}} \frac{\MVA \tan (2 \beta )}{4   (\QHd+\QHu) \l}
   (\QHd-\QHu) \( 2 \la \a \ra \MVA- g_A    v_-^2 \) \nn\\
|m|^2 &\simeq&   g_A \la \a \ra \MVA -  \frac{1}{2} \[ \(g_A^2+g_B^2\)v_-^2 \] \label{eq:lFm_values}
\eea
In Appendix~\ref{app:exactpar} we report the exact formulae.
Thus the only remaining free parameters are $t_\b$ and $\la \a \ra$ and we perform the following analysis of the mass spectrum as a
function of $t_\b$ and $\la \a \ra$.
A lower bound on $\la \a \ra$ as a function of $v_-$ can be obtained, given the approximation~(\ref{vevhierarchyconditioneq}),
from the equation~(\ref{minima3approx})
\be
\la \a \ra \simeq \frac{|m|^2+\(1/2\)\( g_A^2 + g_B^2\) v_-^2}{g_A \MVA}
\ee
where we used the relation
\be
m_{\phi^-}^2=-m_\xi^2=-\la \a \ra g_A \MVA
\ee
Thus the lower bound on $\la \a \ra$ is obtained simply by setting $|m|=0$,
\be
 \la \a \ra_m \simeq \frac{\(1/2\)\( g_A^2 + g_B^2\) v_-^2}{g_A \MVA}
\ee
The condition $\left<\a\right> > \la \a \ra_m$ must hold since otherwise we would have a massless scalar field in the spectrum (see eq. (\ref{eq:phiplusmass})).
Another lower bound, $\la \a \ra_b$, can be obtained from the condition (\ref{eq:destabilizeorigin}),
by solving the minima conditions (\ref{minima1})-(\ref{minima3}) and by substituting the corresponding $\la F_S \ra$, $\l$ and $m$ values (\ref{eq:lFm_values}).
The resulting lower bound can be expressed as
\be
 \la \a \ra > \max\[ \la \a \ra_m, \la \a \ra_b \]
\ee
No upper bound can be imposed, hence we decide to perform our analysis by considering $\la \a \ra \lesssim 100$ TeV.

The parameters $\l$ and $\la F_S \ra$ are of a particular phenomenological importance since they appear in the $\m$ and $b$ terms (see eqs.~(\ref{mueq}) and (\ref{beq})).
In the case A, $\m$ is in the range $(900,6000)$ GeV and $\sqrt b$ is in the range $(50,1200)$ GeV while in the case B, $\m$ is in the range $(500,6000)$ GeV and $\sqrt b$ is in the range $(25,1200)$ GeV. These values are in the right range to solve the $\mu$-problem.

\subsection{Mass spectrum}

\paragraph{A)}
With such choice the gauge vector sector is completely
fixed up to a $t_\b$ dependence. Anyway even such a dependence can be
safely ignored with a very good approximation in the new gauge
sector since the mixing is strongly suppressed. So for each $t_\b$ value
we have
\bea
 M_{Z_1} &\simeq& 10 \TeV \\
 M_{Z_2} &\simeq&  2 \TeV
\eea
where with $Z_1$ we denote the $V_A$-like vector .

\paragraph{B)}
As in the previous case, we just give the $Z_{1,2}$ masses
\bea
 M_{Z_1} &\simeq& 10 \TeV \\
 M_{Z_2} &\simeq& 400 \GeV
\eea
where as in the previous case $Z_1$ is $V_A$-like.

We will not give the exact values of the $Z_0$ mass. It is enough for our purposes to know that they are compatible with the bounds of Section \ref{sec:Charge Bounds}.
Both case A and B are compatible with CDF bounds about $Z'$ direct production \cite{CDF bound}.

\begin{figure}[t]
\begin{center}
\includegraphics[scale=0.3]{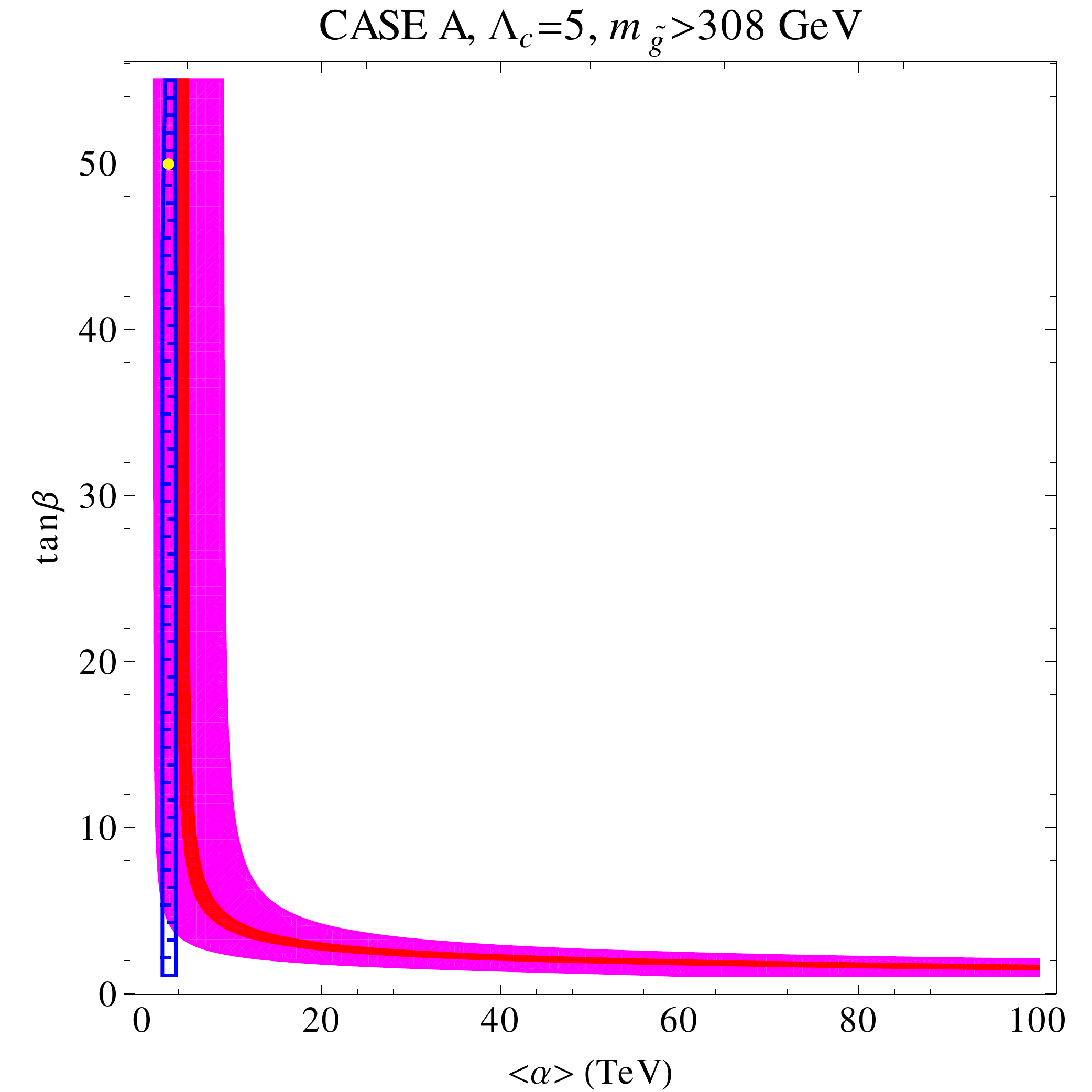}
\includegraphics[scale=0.3]{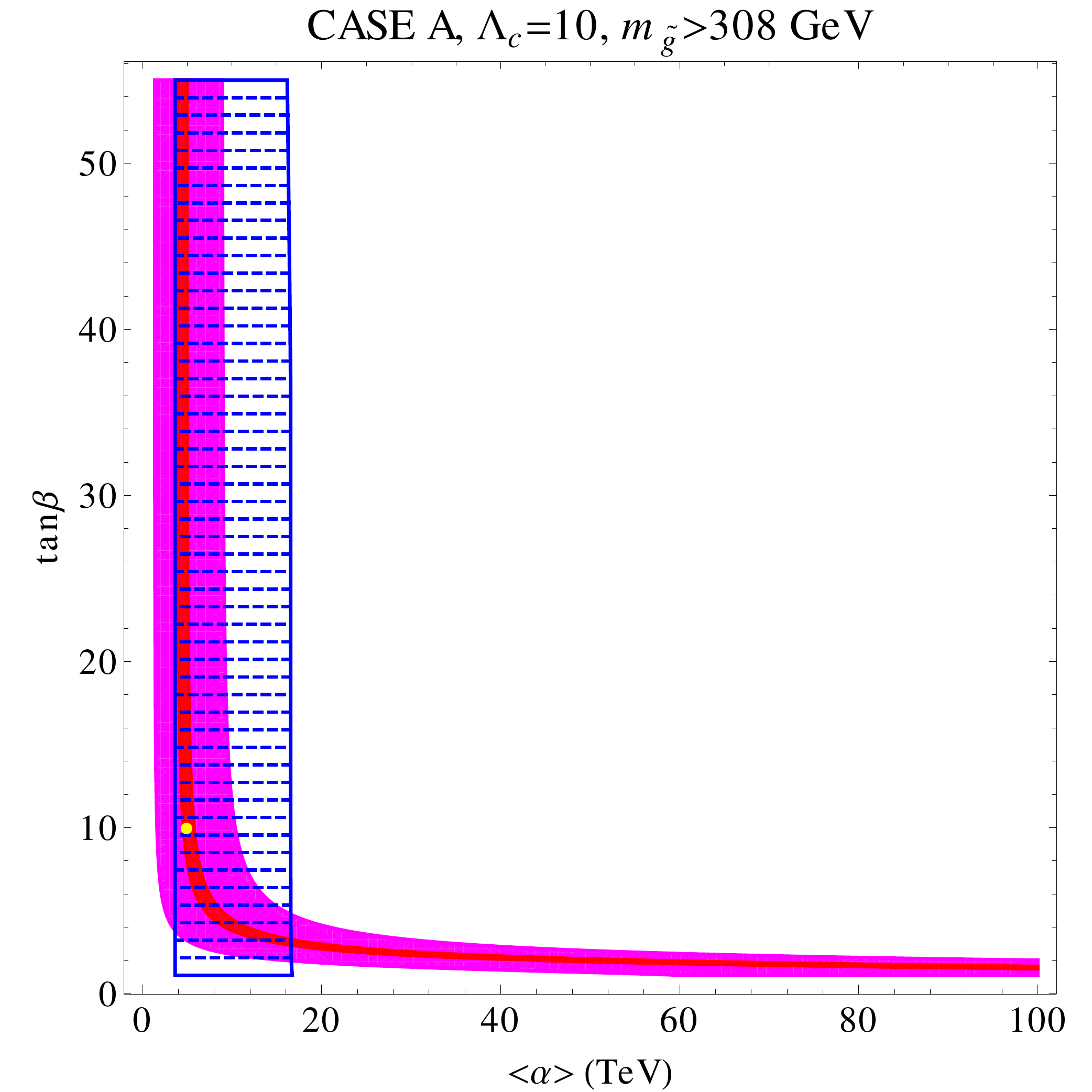}
\includegraphics[scale=0.3]{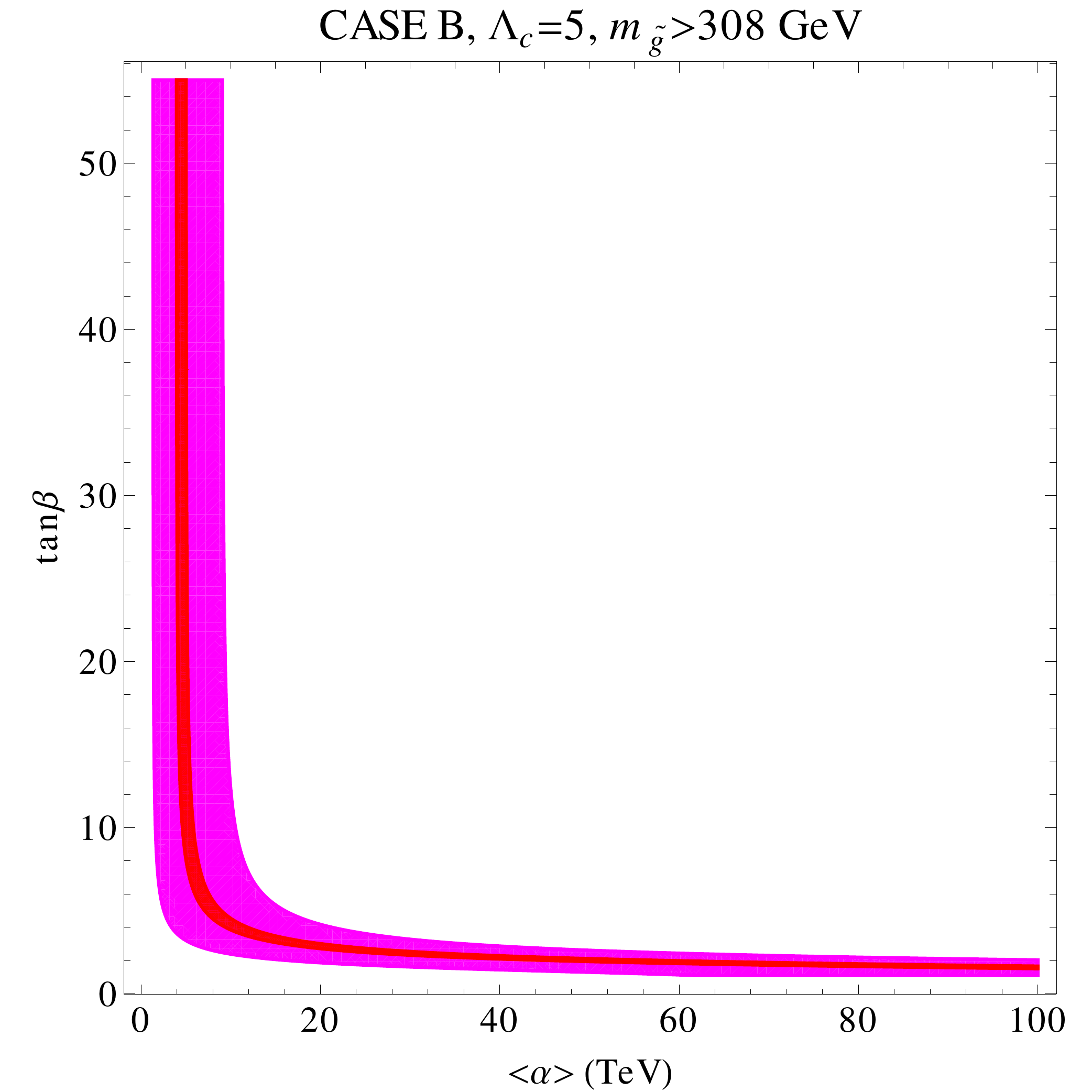}
\includegraphics[scale=0.3]{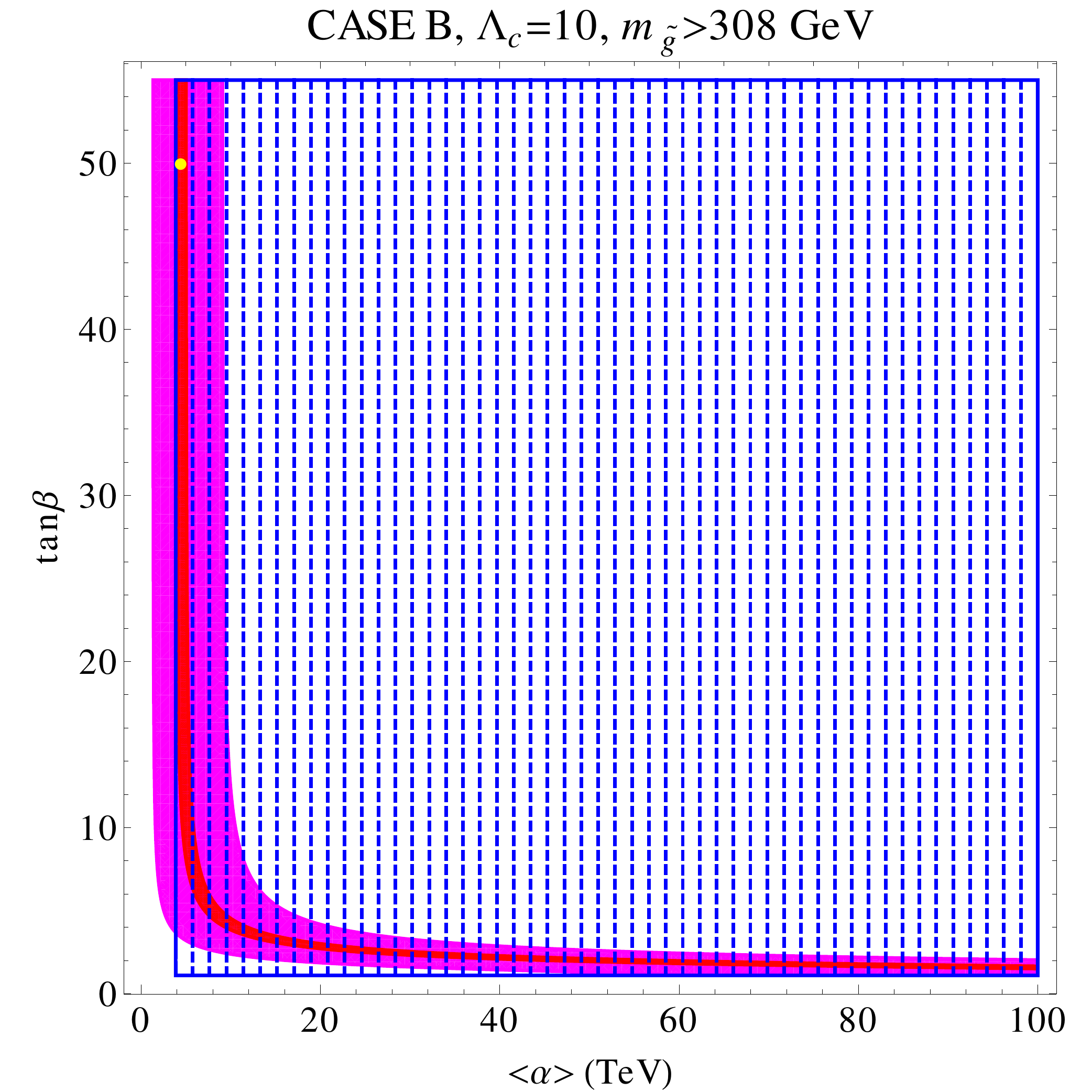}
\end{center}
\vskip -1cm \caption{Allowed $\la \a \ra$ and $\tan\b$ values for case A (up) and case B (down), $\Lambda_c=5$ (left) and $\Lambda_c=10$ (right).
                     The red region is the one in which $\left. M_{h^0}^2 \right|_\text{1-loop} \in [124,126] \GeV$,
                     the magenta region is the one in which $\left. M_{h^0}^2 \right|_\text{1-loop} \in [114.5,131] \GeV$
                     and the blue region satisfies all the mass bounds on the sparticles (from PDG) and requires a neutralino LSP.
                     The yellow dots are our benchmark points.}
\label{fig:alphavstanbcase300}
\end{figure}

\begin{figure}[t]
\begin{center}
\includegraphics[scale=0.3]{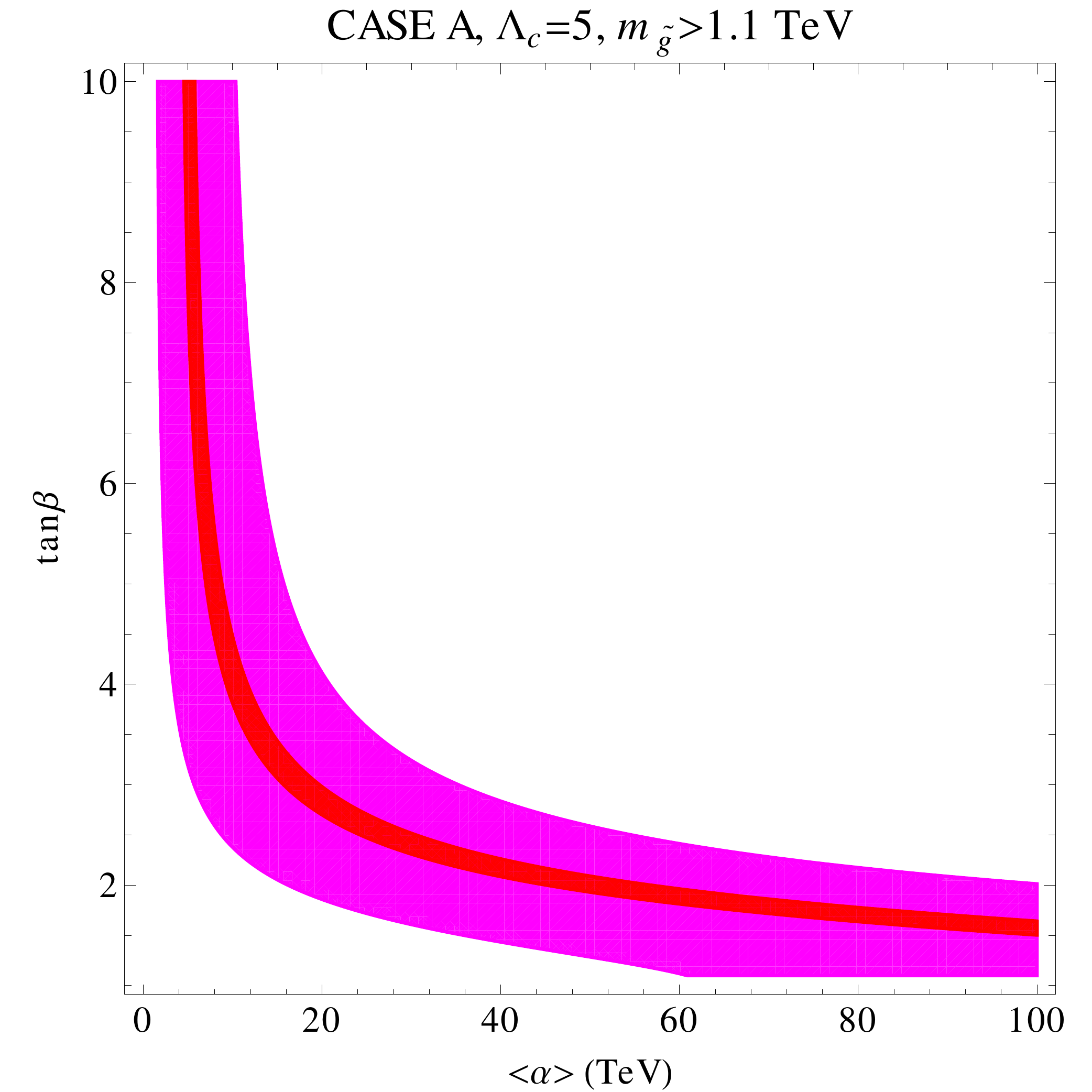}
\includegraphics[scale=0.3]{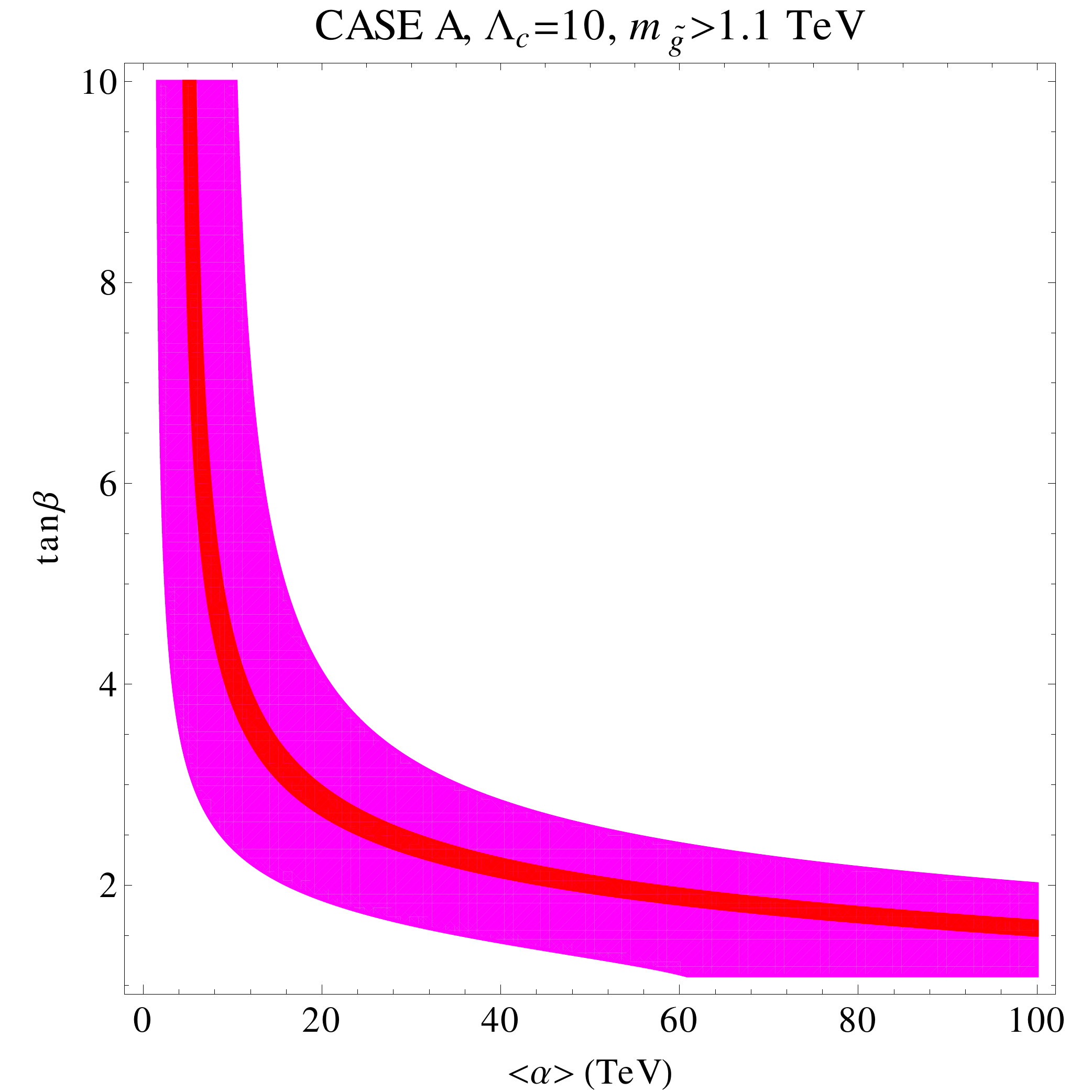}
\includegraphics[scale=0.3]{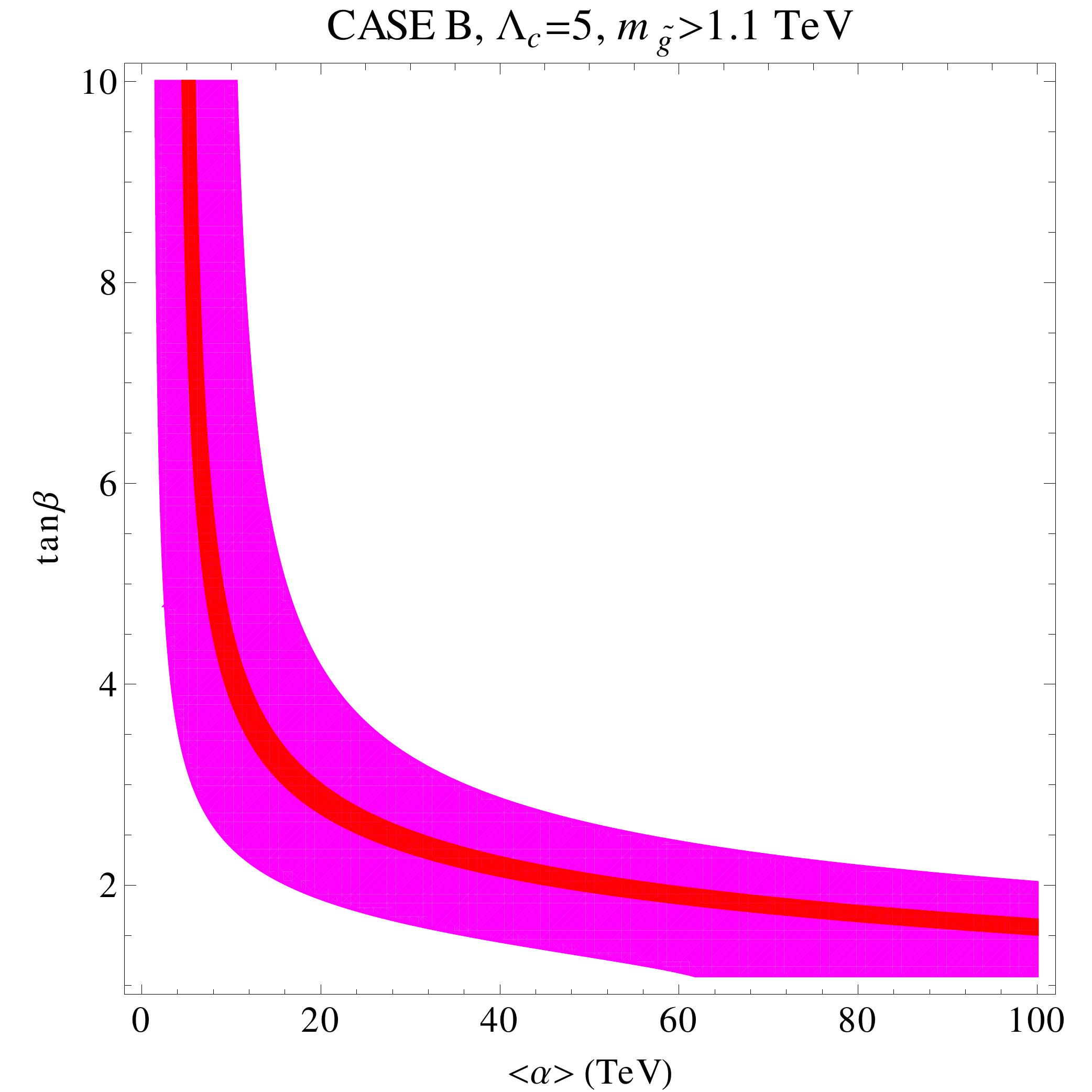}
\includegraphics[scale=0.3]{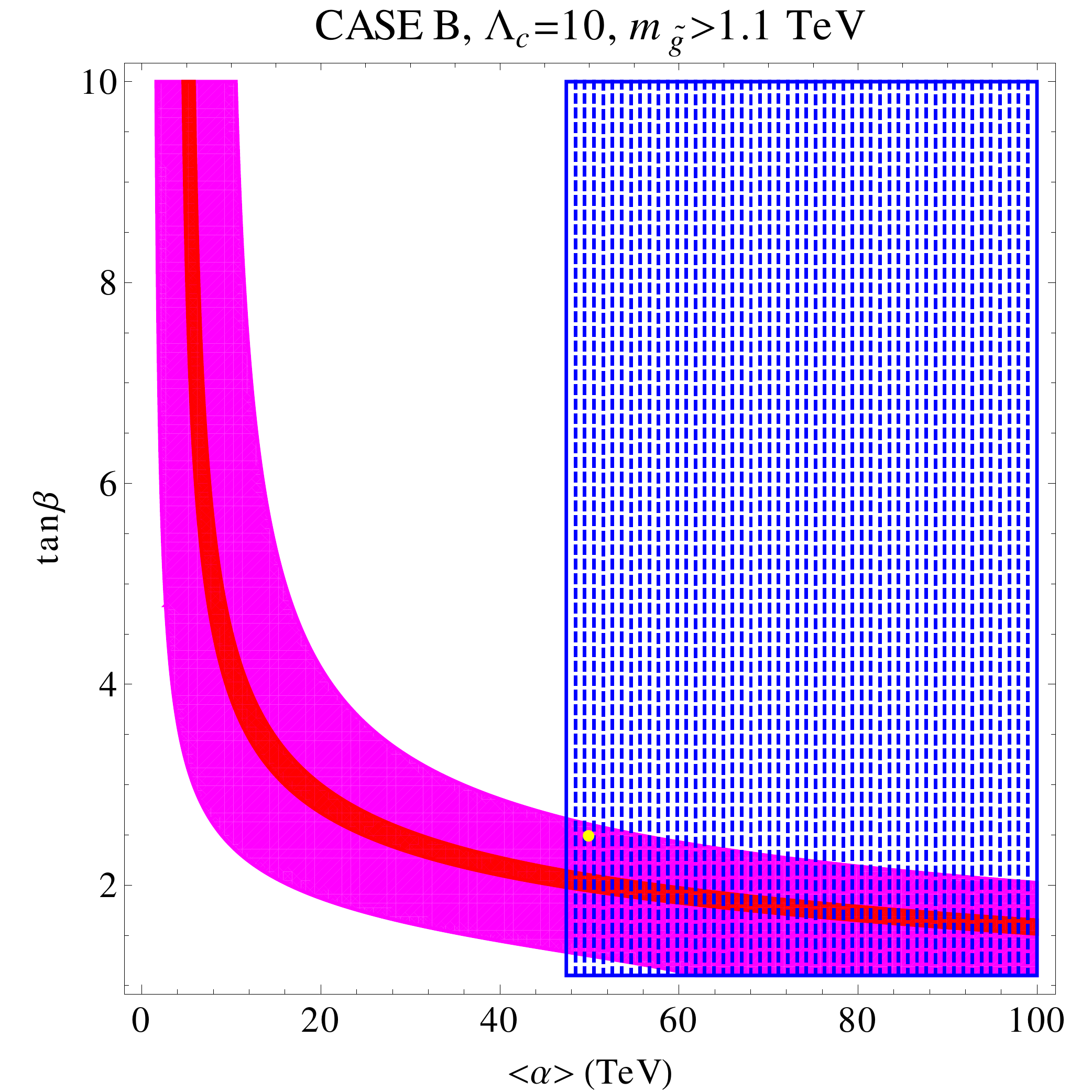}
\end{center}
\vskip -1cm \caption{Allowed $\la \a \ra$ and $\tan\b$ values for case A (up) and case B (down), $\Lambda_c=5$ (left) and $\Lambda_c=10$ (right).
                     The red region is the one in which $\left. M_{h^0}^2 \right|_\text{1-loop} \in [124,126] \GeV$,
                     the magenta region is the one in which $\left. M_{h^0}^2 \right|_\text{1-loop} \in [114.5,131] \GeV$
                     and the blue region satisfies all the mass bounds on the sparticles (from preliminary LHC data) and requires a neutralino LSP.
                     The yellow dots are our benchmark points.}
\label{fig:alphavstanbcase}
\end{figure}

\begin{figure}[p]
\begin{center}
\includegraphics[scale=0.3]{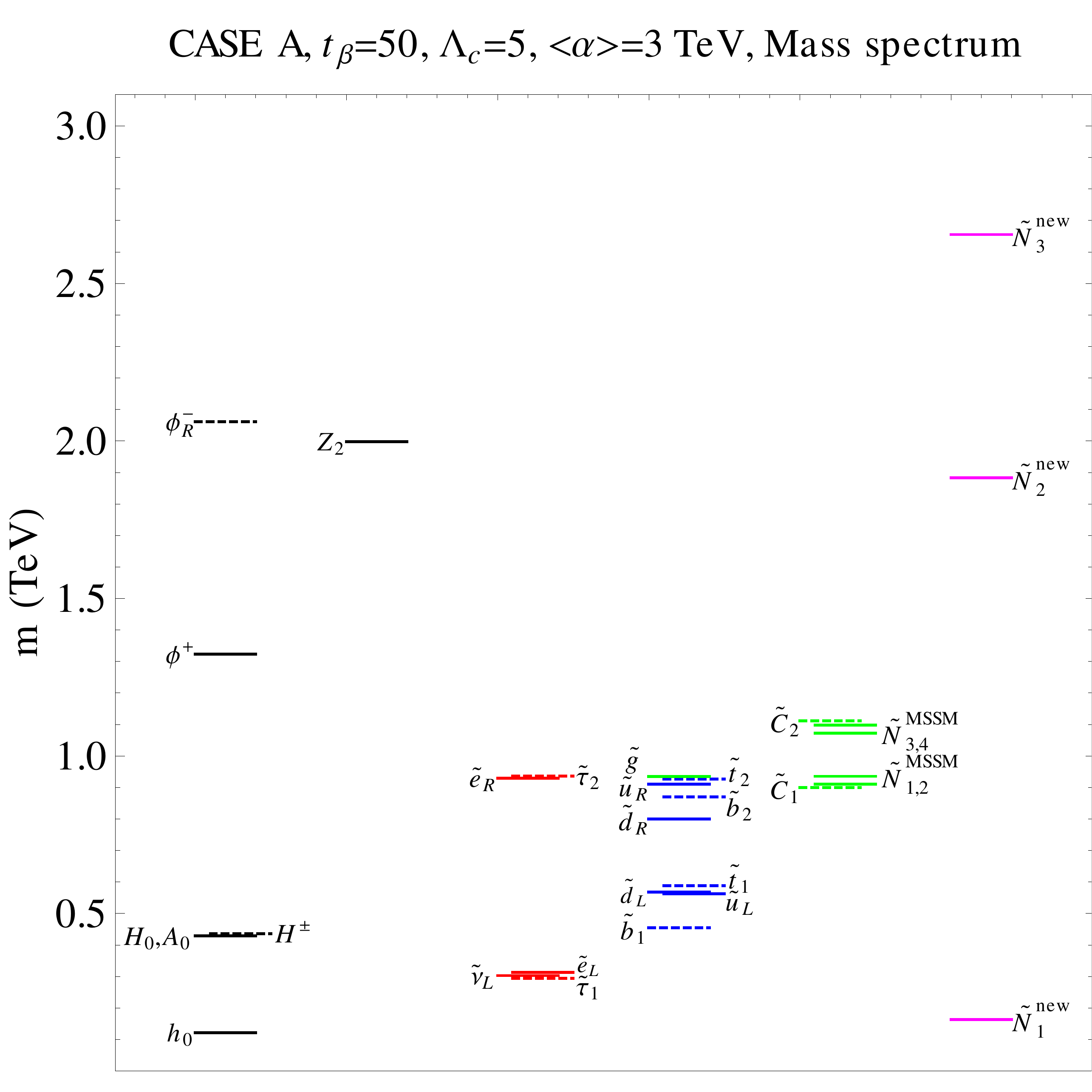}
\includegraphics[scale=0.3]{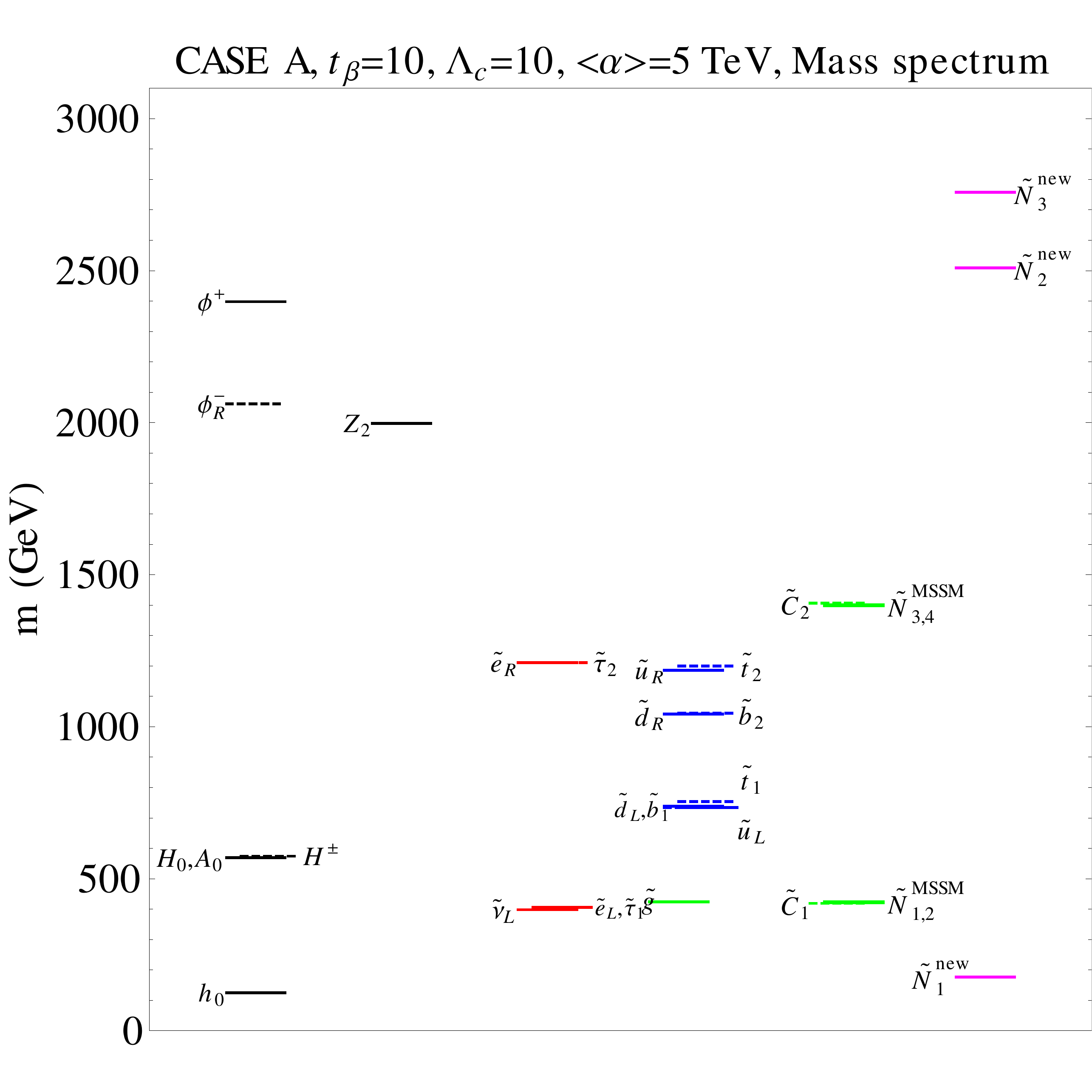}
\end{center}
\vskip -1cm
\caption{Mass spectrum, case A, $\Lambda_c=5$, $\la \a \ra=0.3$ TeV and $t_\b=50$ (left), $\Lambda_c=10$, $\la \a \ra=0.5$ TeV and $t_\b=10$ (right).}
\label{fig:massspectrumA}
\end{figure}

\begin{figure}[p]
\begin{center}
\includegraphics[scale=0.3]{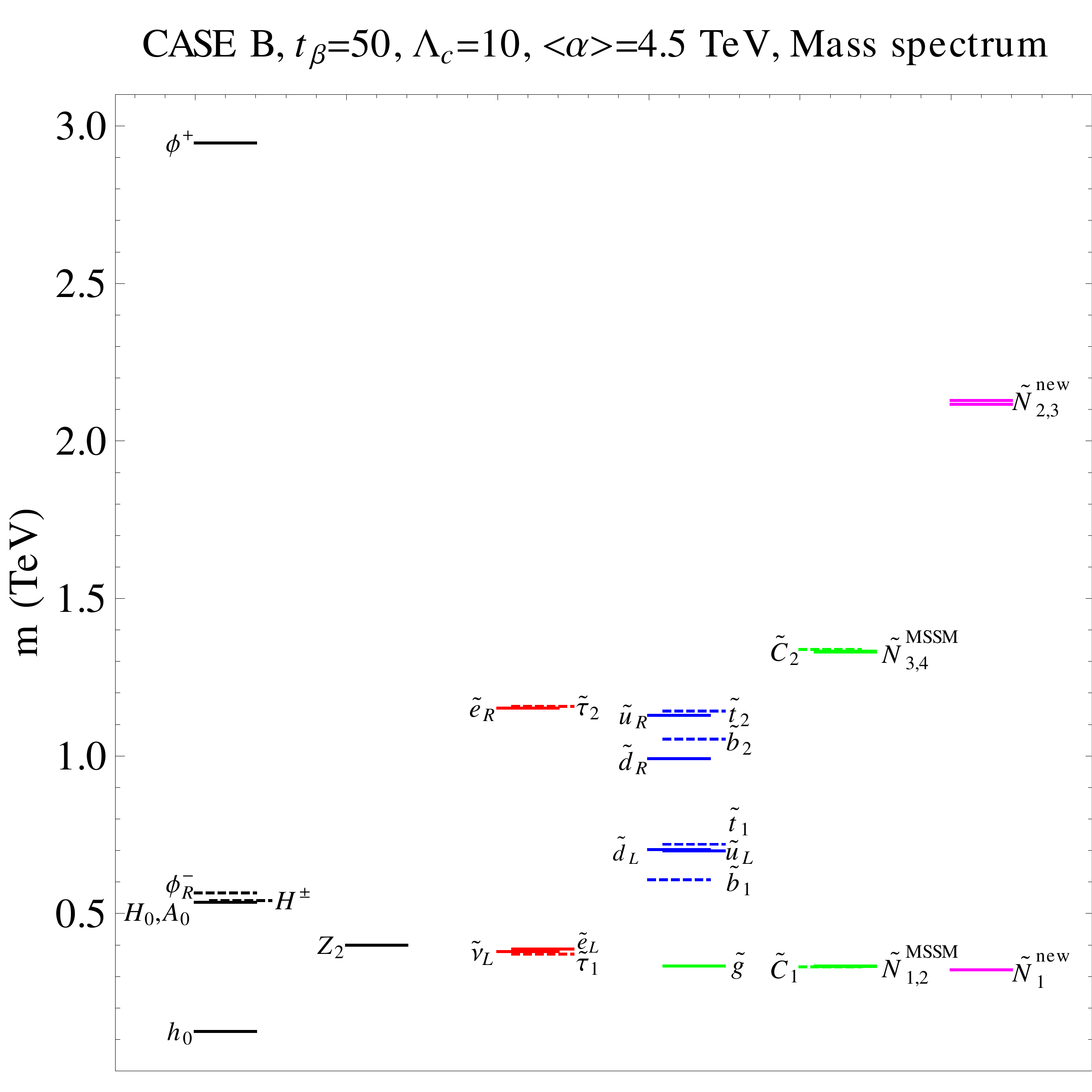}
\includegraphics[scale=0.3]{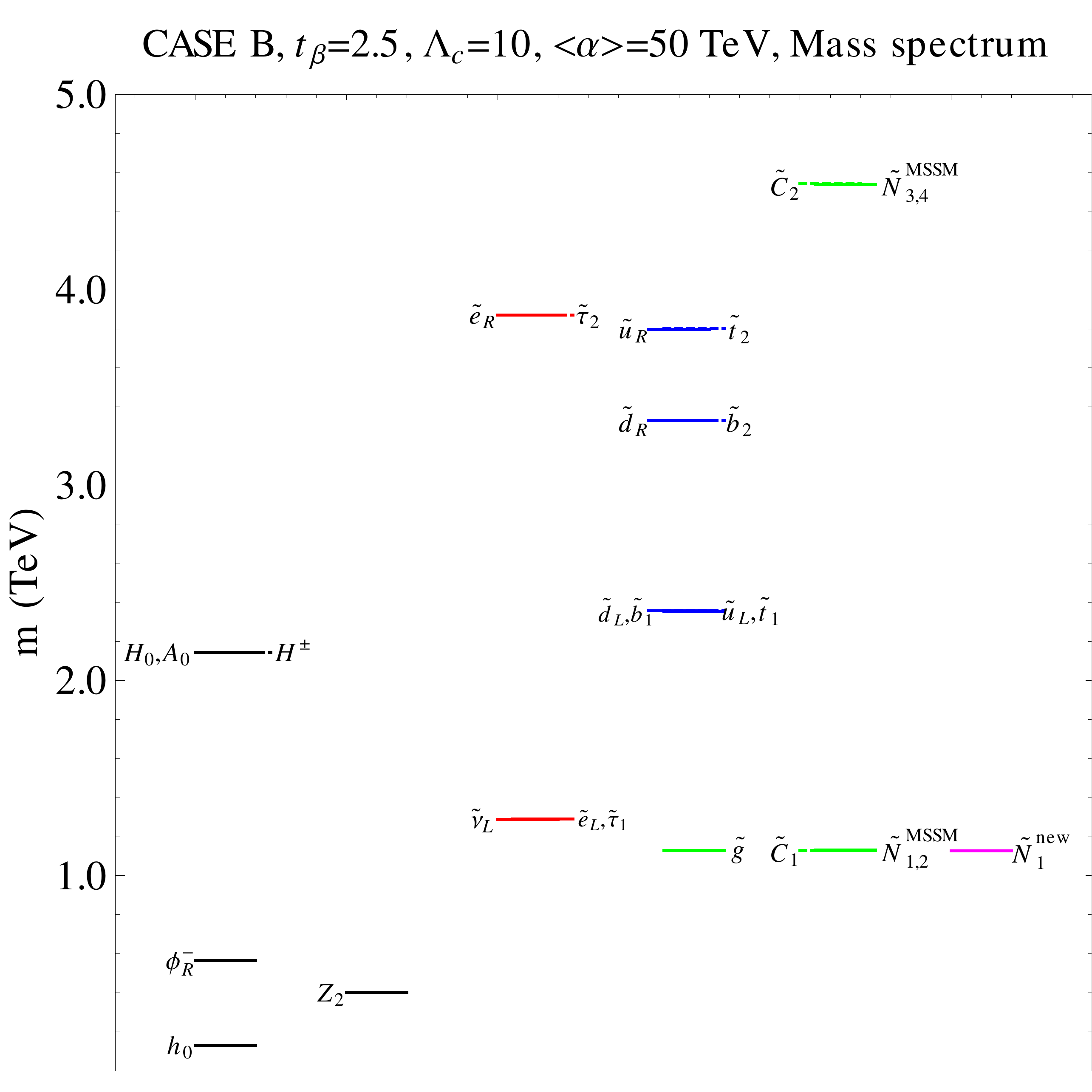}
\end{center}
\vskip -1cm
\caption{Mass spectrum, case B, $\Lambda_c=10$, $\la \a \ra=0.45$ TeV and $t_\b=50$ (left), $\la \a \ra=8$ TeV and $t_\b=2.5$ (right).}
\label{fig:massspectrumB}
\end{figure}

Recent LHC data have restricted the most probable range for the Higgs particle mass to be $[115.5,131]$ GeV (ATLAS) \cite{ATLAS}
and $[114.5,127]$ (CMS) \cite{CMS}. Moreover, there are hints observed by both CMS and ATLAS of an excess of events
that might correspond to decays of a Higgs particle with a mass in a range close to 125 GeV.
So, in Fig.~\ref{fig:alphavstanbcase300} and \ref{fig:alphavstanbcase} we give region plots showing the allowed values of $\la \a \ra$ and $t_\b$ for case A (B) and $\Lambda/\sqrt c=(5) 10 \TeV$.
The red region is the one in which $\left. M_{h^0}^2 \right|_\text{1-loop} \in [124,126] \GeV$ where the $h^0$ mass is computed considering 1-loop corrections.
Since it turns out that the top squarks have small mixing angle and considering the limit $M_{A^0} \gg \MZO$, we have \cite{Martin:1997ns}
\bea
 \left. M_{h^0}^2 \right|_\text{1-loop} &\simeq&
                    \left. M_{h^0}^2 \right|_\text{tree} + \frac{3}{4 \pi^2} s_\b^2 y_t^2 m_t^2 \ln \(m_{\tilde t_1} m_{\tilde t_2} / m_t^2 \) \nn\\
           &\simeq& \left. M_{h^0}^2 \right|_\text{tree} + \frac{3}{2 \pi^2} \frac{m_t^4}{v^2}  \ln \(m_{\tilde t_1} m_{\tilde t_2} / m_t^2 \)
\label{eq:mh0loop}
\eea
where $\left. M_{h^0} \right|_\text{tree}$ is the tree-level $h^0$ mass and we used $m_t = y_t v_u/2= y_t v s_\b/2$.
There is an approximated inverse correlation between $\la \a \ra$ and $t_\b$ in the $h_0$ mass allowed region because the 1-loop correction in
(\ref{eq:mh0loop}) increases for increasing values of $\la \a \ra$ or $t_\b$.
The $h_0$ mass allowed region is almost the same for case A and B because of two reasons
      \begin{itemize}
       \item[i.] the mixing with $\phi^\pm$ is suppressed
       \item[ii.] the parameters $\tilde m_{h_u}, \tilde m_{h_d}, \m, b$ in the scalar potential (\ref{Vhapprox}) are ruled by the square mass parameters
                  $g_A \la \a \ra \MVA$ and $\(g_A v_-\)^2$ and the first one turns out to be dominant.
      \end{itemize}

The magenta region satisfies a milder constraint on the light Higgs boson: $\left. M_{h^0}^2 \right|_\text{1-loop} \in [114.5,131]$ GeV.
In order to be more conservative we imposed the joint constraints of ATLAS and CMS.

The blue region satisfies all the mass bounds on the sparticles and requires a neutralino LSP.
We considered two possibilities: one more optimistic (Fig. \ref{fig:alphavstanbcase300}) using the PDG bounds \cite{Heister:2002jca,:2007ww}
and one more conservative (Fig. \ref{fig:alphavstanbcase}) using recent LHC data \cite{squarkgluino}.
The combination of the gluino mass bound with a neutralino LSP is a strong constraint that reduces drastically the allowed parameter space.
In some cases there is not even a blue region, which means that we cannot satisfy simultaneously all the mass bounds and have a neutralino LSP,
so they are completely ruled out. When the gluino mass bound is from PDG, Case A is allowed, otherwise it is completely ruled out, and
only case B for $\Lambda/\sqrt c=10 \TeV$ presents allowed regions.
We notice that case A favors low $\la \a \ra$ values, while case B favors big $\la \a \ra$ values.
For every allowed case we choose a benchmark point (yellow spots in Fig.~\ref{fig:alphavstanbcase300} and Fig. \ref{fig:alphavstanbcase})

\begin{itemize}
\item[i.]   case A, $\Lambda_c=5$, $\la \a \ra=3$ TeV and $t_\b=50$ so that $\left. M_{h^0} \right|_\text{1-loop} \simeq 121.6 $ GeV
\item[ii.]  case A, $\Lambda_c=10$, $\la \a \ra=5$ TeV and $t_\b=10$  so that $\left. M_{h^0} \right|_\text{1-loop} \simeq 124.7$ GeV.
\item[iii.] case B, $\Lambda_c=10$, $\la \a \ra=4.5$ TeV and $t_\b=50$ so that $\left. M_{h^0} \right|_\text{1-loop} \simeq 125.1 $ GeV.
\item[iv.]  case B, $\Lambda_c=10$, $\la \a \ra=50$ TeV and $t_\b=2.5$ so that $\left. M_{h^0} \right|_\text{1-loop} \simeq 130.1 $ GeV.
\end{itemize}
and we give the full mass spectrum in Fig.~\ref{fig:massspectrumA} and in Fig.~\ref{fig:massspectrumB}.

All the benchmark points share some common features
\begin{itemize}
\item the LSP is the lightest neutralino of the new sector: in case A it is a combination of $\tilde \phi^\pm$ and $\l_B$ while in case B is almost a pure $\l_B$.
\item an approximated mass degeneracy of $H^0$, $A^0$ and $H^\pm$ holds, and their masses satisfy the bounds of~\cite{Schael:2006cr,Heister:2002ev}.
\item the lightest sleptons is a sneutrino, except for $t_\b=50$ when it is $\tilde \t_1$
\item the lightest squark is $\tilde u_L$, except for $t_\b=50$ when it is $\tilde b_1$
\item the first and second family left-handed squarks/sleptons are likely to be lighter than their right-handed counterparts.
      This is at odds with the usual MSSM cases~\cite{Martin:1997ns}.
\item $\tilde C_{1(2)}$ is close in mass with $\tilde N^\text{MSSM}_{1(4)}$.
      $\tilde C_2$ and $\tilde N^\text{MSSM}_{4}$ are heavier than all sfermions.
\item the gluino is close in mass to $\tilde C_1$ and $\tilde N^\text{MSSM}_{1}$ which are gaugino-like.
      Moreover it is lighter than all the squarks except for point i).
      So it turns out to be long lived, specially in case B where the approximated mass degeneracy involves also the LSP.
      Long lived gluinos bind with SM quarks and gluons from the vacuum during the hadronisation process, and produce R-hadrons.
      R-hadrons are among the most interesting searches at LHC. Anyway we will come back to this point with a more detailed study in a forthcoming paper.
\item there is an approximated mass degeneracy between $\tilde e_R$ and $\tilde u_R$ because using the charge constraints~(\ref{QconstK}) and ~(\ref{commonnumeric})
      we get $\QE=3$ and $\QU \simeq 2.9$.
\item $m_{\f^-_R}<m_{\f^+}$ except for point i)
\end{itemize}

Case B points deserve some more comments.

$\f^+$ and $\tilde N^\text{new}_{2,3}$ are out of the plot of point iv) because they are heavier than 6 TeV.
$Z_2$ is among the lightest not SM particle, so it can decay only into SM particles, because of energy and R-parity conservation.
So $Z_2$ is long lived, because SM particles are coupled to $Z_2$ only through the suppressed $V_{A,B}$ mixing or
through the Higgs scalars which present a tiny mixing with $\f^\pm$.

It is not an easy task to compare the resulting spectrum we get for our model with those related to the rich zoology of supersymmetry breaking scenarios.
It is worth to stress anyway that the two representative spectrums showed
in Fig.~\ref{fig:massspectrumB} which encode the key features of our scenarios listed above
are not reproduced in any of the benchmark points showed in~\cite{Allanach:2002nj}.

\section{Conclusions}\label{sec:end}
In this paper we presented a viable mechanism to generate soft supersymmetry breaking terms in the framework of a minimal supersymmetric anomalous extension of the SM.
The crucial ingredient is a non perturbative term in the superpotential~(\ref{newsuperpot}) which couples the St\"uckelberg field $S$ to the Higgs sector.
This term is related to the generation of a suitable $\mu$ and $b$ terms (see Eq.~(\ref{mueq}) and~(\ref{beq})) in the low energy effective action when the St\"uckelberg gets vev.
We argued about the origin of this term from an exotic instanton in an intersecting D-brane setup.
We computed the spectrum of our model as a function of the saxion vev $\left<\a\right>$ and for different choices of the remaining free parameters.
We checked our results against known phenomenological bounds, namely current lower bounds on the mass of the scalar and fermionic superpartners.
We analyzed a scenario in which the anomalous sector is the source of the soft supersymmetry breaking terms while the corresponding vector and
St\"uckelberg multiplets are not present in the low energy effective action.
For what concern the non anomalous sector we took into account two different cases (dubbed case A and case B).

As we stated in Sec.~\ref{sec:pheno}, by applying some phenomenological constraints we were able to derive some bounds on the saxion vev $\la\a\ra$, which is the relevant parameter setting the mass scale of the scalars.
The strongest constraints on $\la \a\ra$ and $t_\b$ comes from the combined requirement of $\left. M_{h^0}^2 \right|_\text{1-loop} \in [124,126] \GeV$
or ($[114.5,131] \GeV$), a neutralino LSP and that all mass bounds (specially the gluino one) are fulfilled.
In Fig.~\ref{fig:alphavstanbcase300} (pre-LHC bounds) and~\ref{fig:alphavstanbcase} (preliminary LHC bounds) we summarize the allowed regions for $\la\a\ra$.
In the first case, by requiring a phenomenological appealing neutralino LSP, we get an allowed $\la\a\ra$ of few TeV up to $10$ TeV for the A and B scenarios respectively.
In the second case (preliminary LHC bounds) we get that only the B scenario is allowed with $\la\a\ra\gtrsim 5$ TeV.
These results can be seen as a bound that a concrete D-brane model has to satisfy.
We deserve this analysis for future work.

In Fig.~\ref{fig:massspectrumB} we explicitly showed two benchmark mass spectrums for our model with $\la \a \ra$ and  $t_\b$
which fulfill the above bounds. The cases shared different peculiar features: the LSP is the lightest neutralino of the new sector, there is a near mass degeneracy between $H^0$, $A^0$ and $H^\pm$,
and between $\tilde e_R$ and $\tilde u_R$,
the lightest sleptons is a sneutrino except for $t_\b=50$ when it is stau, the lightest squark is a $\tilde u_L$ except for $t_\b=50$ when it is a sbottom, the first and second family left-handed squarks/sleptons are typically lighter than their right-handed counterparts.
Moreover in case B the gluino is long lived and can produce R-hadrons.
It turns out that these features are not reproduced in any of the widely studied benchmark points presented in~\cite{Allanach:2002nj}.
\vskip 2cm
\begin{flushleft}
{\large \bf Acknowledgments}
\end{flushleft}

\noindent A.L. acknowledges M.~Bianchi, E.~Kiritsis and R.~Richter for useful discussions and comments.
 A. R. acknowledges M. Raidal for discussions and the ESF JD164 contract for financial support.
\medskip


\appendix

\section{Anomalous Lagrangians} \label{app:ano}
The Lagrangian involved in the anomaly cancellation procedure is
\bea
 \L_S &=& {\frac{1}{4}} \left. \( S + S^\dagger + 2 M_{V_A} V_A \)^2 \right|_{\thth}    \label{Laxion} \\
      && - 2 \left\{ \[\sum_a g_a^2 b^{aa} S \Tr\( W_a W_a \) + g_Y g_A b^{YA} S W_Y W_A \]_{\th^2} +h.c. \right\}
 \nn
\eea
where the index $a=A,B,Y,2,3$ runs over the $U(1)_A$, $U(1)_B$, $U(1)_Y$, $SU(2)$ and $SU(3)$ gauge groups respectively,
and the constants $b^{ab}$ are fixed by the anomaly cancellation.

Since we have only one anomalous $U(1)$ we can avoid the use of GCS terms, distributing the anomalies only on the $U(1)_A$ vertices.
So we have
\bea
 &&     b^{AA} = - \frac{g_A \cA_{AA}}{96\pi^2 M_{V_A}}
 \qquad b^{YY} = - \frac{g_A \cA_{YY}}{32\pi^2 M_{V_A}}
 \qquad b^{22} = - \frac{g_A \cA_{22}}{16\pi^2 M_{V_A}}\nn\\
&&
      b^{33} \ = - \frac{g_A \cA_{33}}{16\pi^2 M_{V_A}}
 \qquad b^{YA} = - \frac{g_A \cA_{YA}}{32\pi^2 M_{V_A}}
\label{bs}
\eea
where the $\cA$'s are the corresponding anomalies
\bea
 \cA_{AA} &=& -10 \QHd^3-9 \QHd^2 (\QL+3 \QQ)-9 \QHd \(\QL^2+3 \QQ^2\) \nn\\
           && -7 \QHu^3-27 \QHu^2  \QQ-27 \QHu \QQ^2+3 \QL^3 \\
 \cA_{YY} &=& -\frac{1}{2} (7 \QHd+7 \QHu+3 \QL+9 \QQ)    \\
 \cA_{22} &=& \frac{1}{2} (\QHd+\QHu+3 \QL+9 \QQ)    \\
 \cA_{33} &=& -\frac{3}{2} (\QHd+\QHu)    \\
 \cA_{YA} &=& 5 \QHd^2+6 \QHd (\QL+\QQ)-\QHu (5 \QHu+12 \QQ)
\eea
where we used the constraints (\ref{QconstW}). Imposing the conditions (\ref{QconstK}) we get
\bea
 \cA_{AA} &=& \frac{1}{64} \(-1168 \QHd^3+1776 \QHd^2 \QHu-996 \QHd \QHu^2+53 \QHu^3\)    \\
 \cA_{YY} &=& -\frac{11}{4} (\QHd+\QHu)    \\
 \cA_{22} &=& -\frac{1}{4} (\QHd+\QHu)     \\
 \cA_{33} &=& -\frac{3}{2} (\QHd+\QHu)    \\
 \cA_{YA} &=& 0 \label{AnoYA}
\eea
We remind that (\ref{AnoYA}) is not a consequence of (\ref{QconstK}), but rather (\ref{QconstK}) is a consequence of imposing (\ref{AnoYA}) in order to cancel
the $U(1)_Y-U(1)_A$ kinetic mixing.

\section{Exact fixed parameters} \label{app:exactpar}
In this Appendix we give the exact values for the $\la F_S \ra$, $\l$, $m$ parameters determined in section \ref{fixpar}.
Solving the minima conditions (\ref{minima1})-(\ref{minima3}), we get
%
\bea
 &&\hspace{-0.8cm} \l^2 = \frac{e^{\,2 \la \a \ra g_A (\QHd+\QHu)/\MVA} }{32} \times \\
                 &&  \times \Big[- g_A \sec (2 \beta )    (\QHd-\QHu) \( 8 \la \a \ra \MVA+    g_A v^2 (\QHd+\QHu)\( \cos (4 \beta )+3 \)-4  g_A v_-^2 \)\nn\\
                 && -8 \la \a \ra     g_A \MVA (\QHd+\QHu)-2 v^2 \( 2  g_A^2    \( \QHd^2+\QHu^2 \)+g_Y^2+g_2^2 \)+4  g_A^2 v_-^2    (\QHd+\QHu)\Big]\nn\\
\quad\nn\\
 &&\hspace{-0.8cm} \la F_S \ra = - e^{\la \a \ra g_A (\QHd+\QHu)/\MVA}\times \frac{\MVA \tan (2 \beta )}{8  g_A   (\QHd+\QHu) \l}  \times\nn\\
&&\times \Big[  g_A (\QHd-\QHu) \( 4 \la \a \ra \MVA+ g_A v^2   (\QHd+\QHu)-2  g_A v_-^2 \)+\nn\\
&&v^2 \cos (2 \beta ) \(  g_A^2    (\QHd-\QHu)^2+g_Y^2+g_2^2 \) \Big] \nn\\
\quad\nn\\
&&\hspace{-0.8cm}
|m|^2=  g_A \la \a \ra \MVA - \frac{1}{2} \[  \(g_A^2+g_B^2\)v_-^2 + g_A^2 v^2 \( \QHd c_\b^2+ \QHu s_\b^2 \)  \] \nn
\eea

\end{document}